\documentclass[
 reprint,
 showkeys,
 superscriptaddress,
 amsmath,amssymb,
 aps,
floatfix,
]{revtex4-2}
\usepackage{graphicx}
\usepackage{dcolumn}
\usepackage{bm}
\usepackage{hyperref}
\usepackage{siunitx}
\begin{document}

\preprint{APS/123-QED}

\title{Shape switching and tunable oscillations of adaptive droplets}

\author{Tim Dullweber}
    \affiliation{
European Molecular Biology Laboratory, Heidelberg, Germany\\
}
    \affiliation{
Department of Physics and Astronomy, Heidelberg University, Heidelberg, Germany
}
\author{Roman Belousov}
     \affiliation{
European Molecular Biology Laboratory, Heidelberg, Germany\\
}
\author{Camilla Autorino}
     \affiliation{
European Molecular Biology Laboratory, Heidelberg, Germany\\
}
    \affiliation{Faculty of Biosciences, Heidelberg University, Heidelberg, Germany}
\author{Nicoletta Petridou}
     \affiliation{
European Molecular Biology Laboratory, Heidelberg, Germany\\
}
\author{Anna Erzberger}
 \email{erzberge@embl.de}
    \affiliation{
European Molecular Biology Laboratory, Heidelberg, Germany\\
}
    \affiliation{
Department of Physics and Astronomy, Heidelberg University, Heidelberg, Germany
}

\date{\today}

\begin{abstract}
Living materials adapt their shape to signals from the environment, yet the impact of shape changes on signal processing and associated feedback dynamics remain unclear. We find that droplets with signal-responsive interfacial tensions exhibit shape bistability, excitable dynamics, and oscillations. The underlying critical points reveal novel mechanisms for physical signal processing through shape adaptation in soft active materials. We recover signatures of one such critical point in experimental data from zebrafish embryos, where it supports boundary formation.
\end{abstract}

\maketitle

Dynamic and non-trivial geometries are hallmarks of soft active matter, because intrinsic stress fields drive autonomous shape changes in deformable materials \cite{mietke2019self, salbreux2017mechanics, marchetti2013hydrodynamics, shankar2022topological, binysh2022active}. In turn, boundary geometry influences stresses and material properties, or determine how macroscopic work is extracted from microscopic activity \cite{ray2023rectified, araujo2023steering}. In complex materials that process signals and adaptively respond to their environment \cite{mcevoy2015materials,zhao2023chemotactic, ziepke2022multi, gonzales2023bidirectional}, geometry-dependent feedback arises when signal processing depends on the system's shape.

In particular living materials possess internal degrees of freedom that adjust their mechanical properties in response to peripheral signals. In cells, signals trigger biochemical processes including the regulation of genes that control the molecular composition in the bulk and at the surface \cite[Chap.~15]{alberts2022molecular}. Inhibitory signaling interactions between neighboring cells for example lead to the spontaneous symmetry-breaking of such internal states, giving rise to distinctly shaped cell types \cite{pisarchik2022multistability, sjoqvist2019say}. The resulting mechanochemical feedback governs the spatial organisation of diverse multicellular systems 
\cite{dullweber2023mechanochemical, sprinzak2021biophysics, erzberger2020mechanochemical}, enabling them to solve tasks including locomotion \cite{luo2023autonomous,shah2021soft, venturini2020nucleus, bodor2020cell}, self-healing \cite{terryn2021review, ajeti2019wound}, and the self-organisation of complex structures \cite{sun2023mean,gov2018guided,paluch2009biology}. 
Yet, how internal cellular states interact with shape dynamics is an open question~\cite{linding2021shapes, prasad2019cell, paluch2009biology}, and more generally, how the phase space of adaptive shape-changing materials depends on geometry is unclear.

Uncovering theoretical principles that govern the rich physics of adaptive active matter systems requires minimal, tractable paradigms. Here, we consider adaptive droplets that change their surface tension in response to signals received at contacts with other droplets. To analyse how nonlinear signal processing interplays with geometrical nonlinearities, we introduce a minimal set of equations governing the macroscopic droplet states, which can be derived from microscopic reaction-diffusion dynamics of signaling and adhesion molecules, as shown in companion paper \cite{PREJoint}. 

We show that positive mechanochemical feedback gives rise to multiple stable droplet configurations, while mutually inhibitory interactions drive excitability and oscillations of droplet shapes. %These regimes arise from a saddle-node pitchfork codimension-2 bifurcation point.

By applying our theory to data from zebrafish embryos, we show how feedback between cell-cell adhesion and signaling supports the formation of distinct tissue regions, highlighting the regulatory role of mechanochemical interactions in developmental patterning.

\begin{figure}[b]
    \centering
    \includegraphics[width=86mm]{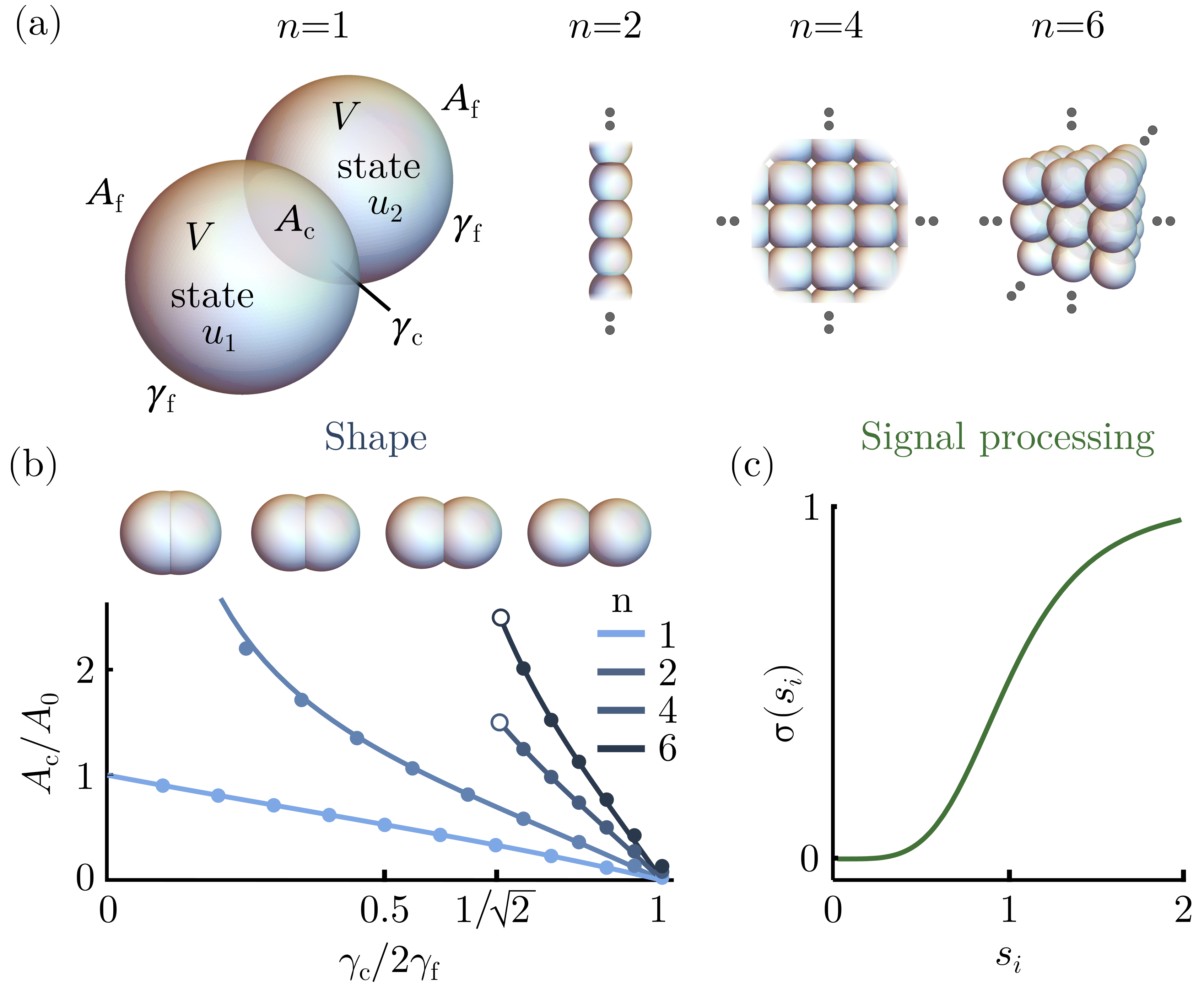}
    \caption{
    Droplets with equal conserved volumes and uniform surface tensions $\gamma_{\rm{f}}$ and $\gamma_{\rm{c}}$ in configurations with $n$ contacts and no triple or higher-order junctions form equilibrium configurations in which the total contact area per droplet (b) depends on $n$ and the ratio $\gamma_{\rm{c}}/2\gamma_{\rm{f}}$~[Eq.~\eqref{eq:contact_area}] (points: numerical results~\cite{brakke1992surface}, empty circles: appearance of higher-order junctions, images for $\gamma_{\rm{c}}/2\gamma_{\rm{f}}=\{0.2,0.4,0.6,0.8\}$). (c) The internal states $u_{i=1,2,...}$  determine $\gamma_{\rm{c}}$ and respond sigmoidally to $A_{\rm{c}}$ dependent signals $s_i$.}
    \label{fig:Fig1}
\end{figure}

\paragraph*{Adaptive Young-Laplace droplets.---}We consider configurations of Young-Laplace droplets with interfacial areas governed by the conjugate uniform surface tensions at fixed volumes [Fig.~\ref{fig:Fig1}(a)]. The total surface energy in a system of $N$ droplets is
\begin{equation}
	E = \sum_{i=1}^{N} \frac{\gamma_{\rm{c}}}{2} A_{\mathrm{c},i}
	+ \gamma_{\rm{f}} A_{\mathrm{f},i},
    \label{eq:surface_energy}
\end{equation}
in which $\gamma_{\rm{c}}$ and $\gamma_{\rm{f}}$ are the surface tensions of the contact interfaces $A_{\mathrm{c},i}$ and the free surface areas $A_{\mathrm{f},i}$, respectively. 

For configurations in which each droplet has $n$ neighbors and no triple or higher-order junctions are present, the total contact area per droplet that minimizes the surface energy Eq.~\ref{eq:surface_energy} is
\begin{widetext}
\begin{align}
    \dfrac{A_{\rm{c}}}{A_0} = 
    n \left[ 1 - \left(\frac{\gamma_{\rm{c}}}{2\gamma_{\rm{f}}}\right)^2 \right]
    \left[ 
        \dfrac{2}
        {\left(2 - \dfrac{\gamma_{\rm{c}}}{2\gamma_{\rm{f}}}\right)
        \left(1 + \dfrac{\gamma_{\rm{c}}}{2\gamma_{\rm{f}}}\right)^2
        - (n-1) \left(2 + \dfrac{\gamma_{\rm{c}}}{2\gamma_{\rm{f}}}\right)
        \left(1 - \dfrac{\gamma_{\rm{c}}}{2\gamma_{\rm{f}}}\right)^2}
    \right]^{\frac{2}{3}},
    \label{eq:contact_area}
\end{align}
\end{widetext}
in which the reference area $A_0 = (3V/2)^{2/3}\pi^{1/3}$ is defined by the conserved droplet volume $V$ (\cite{sm}, Sec.~\ref{sec:wet_foams}). While Eq.~\ref{eq:contact_area} holds for doublets within the full stable-contact regime $0\leq \gamma_{\rm{c}} \leq 2 \gamma_{\rm{f}}$, square ($n=4$) and cubic ($n=6$) lattices form higher order junctions when $\gamma_{\rm{c}}/2\gamma_{\rm{f}} \leq 1/\sqrt{2}$ [Fig.~\ref{fig:Fig1}(b)].

We take the adhesion between droplets to depend on the signals received at their contact surfaces. Therefore, to each droplet $i \in {1,2,...,N}$ we assign a dimensionless \emph{internal state} $u_i \in [0, 1]$---a variable which increases in response to received signals in a saturating manner~[Fig.~\ref{fig:Fig1}(c)]~\cite{herszterg2023signalling,erzberger2020mechanochemical,corson2017self}. When the internal states drive active processes that increase the adhesion between droplets, a bilinear relation that links the tension at the interface between droplets $i$ and $j$ to their internal states is obtained when expanding around a constant $\gamma_0$ (\cite{PREJoint})
\begin{align}
    \gamma_{\rm{c}} = \gamma_0 - \gamma_{\rm{A}} u_i u_j,
    \label{eq:tension_regulation}
\end{align}
in which the second term contains the active contributions in response to signaling, and $\gamma_0$ all state-independent contributions. 
Diverse microscopic processes can modulate the effective tension at the surface of cells, including the generation of stresses by molecular motors~\cite{sitarska2020pay, chugh2017actin, nambiar2009control}, and biochemical regulation of adhesion \cite[Chap.~19]{alberts2022molecular, schwarz2013physics, maitre2012adhesion}. Equation~\eqref{eq:tension_regulation} can be derived from mass-action reaction kinetics, in which the internal states control the active production of adhesion molecules in the bulk, which bind across the interface and lower the surface energy ~\cite{PREJoint}.
The \emph{adaptive adhesion coefficient} $\gamma_{\rm{A}}$ captures the coupling strength between internal states and interfacial tension. Because adhesion reduces the tension, the adaptive term is always negative and the coefficient can be expressed as $\gamma_{\rm{A}} = \epsilon/\lambda^2$ in terms of an energy $\epsilon$ per adhesion complex and a length scale $\lambda$ dependent on the rate of the adhesion molecules' turnover at the bulk-surface interface \cite{PREJoint}. The adaptive tension vanishes without signaling and it always increases the interfacial area $A_{\rm{c}}(\gamma_0) \leq A_{\rm{c}}(\gamma_{\rm{c}})$  [Eq.~\eqref{eq:contact_area}] up to the saturation limit of the internal states.

The state of a droplet $u_i$ is taken to evolve according to a generic signaling interaction~\cite{erzberger2020mechanochemical, corson2017self}
\begin{align}
    \tau_{\rm{u}} \frac{du_i}{dt} = \sigma(s_{i}) - u_i,
    \label{eq:u}
\end{align}
in which $\sigma(s_{i})$ is an increasing sigmoidal response function to a signal $s_{i}$ received by droplet $i$, which we model with a Hill function $\sigma(s_{i}) = s_{i}^h/(s_{i}^h+1)$~\cite{collier1996pattern, binshtok2018modeling} [Fig.~\ref{fig:Fig1}(c)]. We use in the following a first-order relation between the received signal $s_i$ and the available external signal $\phi$, 
\begin{equation}
	s_{i} =  \chi \frac{A_{\rm{c}}}{A_0} \phi, 
    \label{eq:signal}
\end{equation}
which takes into account a linear dependence of the received signal on the contact area~\cite{khait2016quantitative,shaya2017cell} with a \emph{susceptibility} $\chi$. Equations~\eqref{eq:u}--\eqref{eq:signal} can be derived from reaction-diffusion dynamics of biochemical signaling molecules where $u_i$ is defined as a rescaled and normalized concentration of a regulator molecule, such as a transcription factor \cite{PREJoint}. These microscopic equations relate $\chi$ to the steady-state concentrations and kinetic rates associated with the turnover, binding, and processing of signaling molecules in a cell. The linear dependence of transmitted molecular signals on the contact area is valid when diffusion across the contact line and the loss of molecules through biochemical processes at the surface are negligible. 

Provided that the adaptive droplets relax to their equilibrium configuration fast compared to the signaling time scale $\tau_{\rm{u}}$, the collective droplet dynamics are governed by Eqs.~\eqref{eq:contact_area}--\eqref{eq:signal}. Indeed, for cells and cellular aggregates, the viscoelastic response to mechanical stresses is usually on the order of seconds to minutes \cite{wyatt2016question, tran1991time}, while protein production and degradation take place on timescales of tens of minutes to several hours \cite{buccitelli2020mrnas, SHAMIR20161302}. However, surface-mechanical changes due to local biochemical processes are faster \cite{SHAMIR20161302}, and could lead to interplay with the shape dynamics depending on the frictional timescale.
In the low-friction limit, two coupling coefficients characterize the feedback between the system's geometry and the internal states: the adaptive adhesion coefficient $\gamma_{\rm{A}}$ controls the effect of signaling on the surface mechanics and thereby shape, and the signal susceptibility $\chi$ captures how transmitted signals depend on the contact area.

\paragraph*{Tension adaptation produces a shape transition}

\begin{figure}[ht!]
    \centering
    \includegraphics[width=8.6cm]{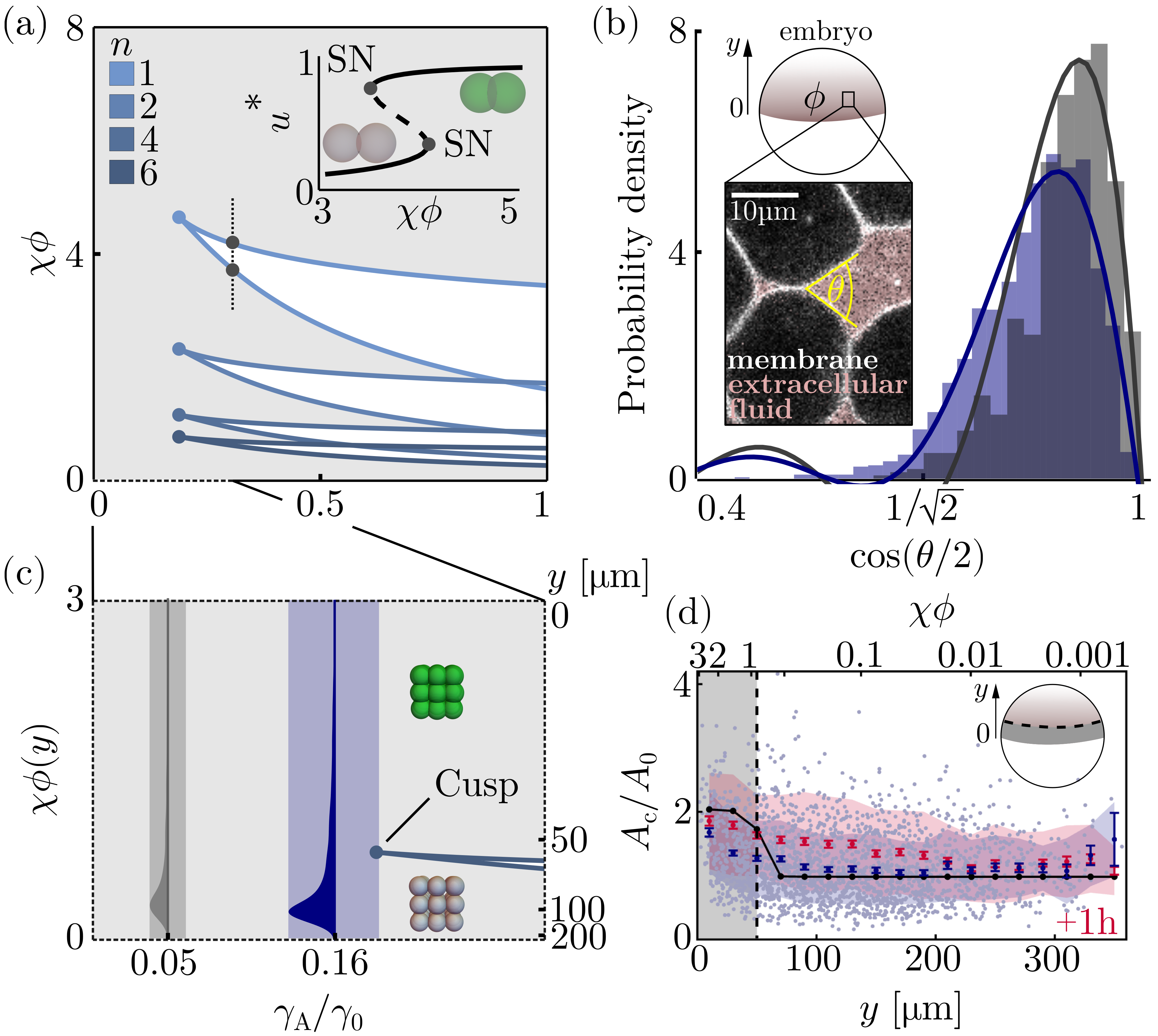}
    \caption{Adaptive tension produces a shape transition which supports patterning in zebrafish embryos. (a) Positive feedback between contact-dependent signals and adaptive adhesion produces bistability between small- and large-contact configurations (inset: bifurcation curve along dotted line, $n$: number of neighbors, blue points: cusps).
    (b) Cell-cell contact angle measurements $\theta$ from fluorescence microscopy images (inset) of blastoderm in unperturbed wildtype zebrafish embryos (blue, 2132 cells from 5 embryos) and \emph{silberblick} mutants with disrupted adhesion regulation (gray, 806 cells from 3 embryos) allow the estimation of parameters $\gamma_0/2\gamma_{\rm{f}}$, $\gamma_{\rm{A}}/\gamma_0$, and $\chi \phi_0$ using simulation-based inference (\cite{Tejero-Cantero2020} and \cite{sm}, Sec.\,\ref{sec:sbi}) (solid lines: best fit).%, assuming a non-uniform signaling field $\langle \phi(y) \rangle~=~\phi_0 \exp(-y/\SI{40}{\micro\metre})$ \cite{muller2012differential}. 
    (c) The inferred parameter distributions locate wildtype embryos close to the cusp, whereas mutants lose adaptive adhesion (shaded regions: standard error from cross-validation (\cite{sm}, Sec.~\ref{sec:sbi}), images for $\gamma_{\rm{c}}/2\gamma_{\rm{f}}=\{0.87,0.71\}$). (d) Mapping external signal levels to spatial positions (also right axis in (c)), we predict---without further fitting---a switch from high- to low-contact configurations at $\sim$ \SIrange{50}{60}{\micro\metre} above the tissue margin (inset), matching the size of the subsequently forming rigid tissue (dark blue and red: experimental mean and standard error at consecutive timepoints, shaded areas: standard deviations, dots: individual data points, black: theoretical profile for inferred $\gamma_{\rm{A}}/\gamma_0=0.16$).  
    (a)--(d) were computed with Eq.~\eqref{eq:contact_area} linearized around $\gamma_{\rm{c}}/2\gamma_{\rm{f}}=1$~\cite[Fig.~\ref{fig:wet_foam_contacts}(d)]{sm}).}
    \label{fig:embryo}
\end{figure}

For constant external signals $\phi$, numerical continuation (\cite{sm}, Sec.~\ref{sec:numerical_methods}) of Eqs.~\eqref{eq:contact_area}--\eqref{eq:signal} with varying coupling parameters $\gamma_{\rm{A}}$ and $\chi$ reveals a bistable regime, delineated by two saddle-node (SN) lines originating from a codimension-2 cusp point [Fig.~\ref{fig:embryo}(a)]. Within this regime, two stable droplet configurations coexist: a weakly adhesive configuration with small interfaces and uniform low levels of $u_i$, and a large-contact configuration with high $u_i$.  This positive feedback effect emerges because the adaptive adhesion increases the contact areas and thereby amplifies the received signals in the droplets. Since the contact area limits the amplitude of adhesion-inducing signals, small contact areas remain weakly adhesive, whereas the large ones support transmission of strong signals, in their turn promoting robust adhesion.

We expect this shape transition to occur in diverse systems in which area-dependent signals affect mechanical changes. As an example of a specific mechanochemically regulated system, we experimentally investigated cell shapes in the zebrafish blastoderm. In this embryonic tissue, it was previously shown that Nodal, an extracellular signaling molecule involved in cell fate specification \cite{hill2022establishment,muller2012differential}, increases intercellular adhesion \cite{petridou2021rigidity}, and that cell-cell contacts can in turn enhance the competence of cells to respond to Nodal \cite{barone2017effective}. Moreover, the external level of Nodal varies spatially, decreasing from the tissue margin to the embryo pole, thus allowing us to test if the cells undergo the predicted switch from strong to weak adhesion as a function of the external signal in Eq.~\eqref{eq:signal}, which thereby acts as the control parameter. While the structure of the zebrafish blastoderm resembles a disordered wet-limit foam with an average of six contacts per cell \cite{drenckhan2015structure}, we model it for simplicity as an ordered lattice with $n = 6$ contacts (cubic) in the small-angle limit, consistent with the data (\cite[Fig.~\ref{fig:wet_foam_contacts}(d-e)]{sm}). Given the typically low concentrations of signaling molecules~\cite{rogers2011morphogen, gregor2007probing}, we model fluctuations of the local level of Nodal by a Gamma distribution, whose mean follows an exponentially decaying profile $\langle\phi(y)\rangle = \phi_0 \exp(-y/\xi)$ from its source at the tissue margin with a characteristic length of $\xi = \SI{40}{\micro\metre}$~\cite{muller2012differential}. The variance of the Nodal level $\sigma_{\phi}^2 = \langle \phi \rangle$ is motivated by the Poissonian statistics of density fluctuations~\cite[Appendix\,III]{chandrasekhar1943stochastic}.
To test our predictions, we measured the distribution of cell-cell contact angles $\theta$ from fluorescence microscopy images of embryos taken five hours post fertilization at different positions $y_j\geq0$ along the embryo axis [Fig.~\ref{fig:embryo}(b)], and evaluated Eqs.~\eqref{eq:contact_area}--\eqref{eq:signal} at each $y_j$ in the local approximation, i.e. neglecting spatial variations of Nodal across nearest-neighbor cells and any non-local effects of area coupling. Using simulation-based inference~\cite{Tejero-Cantero2020} we then estimated the three unknown parameters from the samples of $\theta$ [Fig.~\ref{fig:embryo}(b)], obtaining $\gamma_0/2\gamma_{\rm{f}} = 0.87\pm0.01, \gamma_{\rm{A}}/\gamma_0=0.16\pm0.03$ and $\chi \phi_0=3.1\pm0.8$ (standard error from cross-validation, see \cite{sm} Sec.~\ref{sec:contact_angle_modeling}). %The small value of the ratio $\gamma_{\rm{A}}/\gamma_0$ is consistent with Eq.~\eqref{eq:tension_regulation} being a lowest-order expansion around a constant. 
Furthermore, genetic perturbation of adhesion regulation in \emph{silberblick} mutants \cite{petridou2021rigidity, witzel2006wnt11, ulrich2005wnt11} yielded a significant reduction in the adaptive tension coefficient as expected ($\gamma_{\rm{A,SLB}}/\gamma_0=0.05\pm0.01$, inferred with the two other parameters $\gamma_0/2\gamma_{\rm{f}}$ and $\chi\phi_0$ left unaltered [Fig.~\ref{fig:embryo}{c}]).

Overall, we find that the estimated parameters of blastoderm cells are close to the critical cusp point of the bistable regime, which locates the transition between low- and high-contact regimes at approximately \SIrange{50}{60}{\micro\metre} above the margin of the tissue [Fig.~\ref{fig:embryo}(d), obtained without further fitting]. This length corresponds to the size of the subsequently developing rigid tissue region, which will later form the internal parts of the organism \cite{van2015temporal}.

\paragraph*{Adaptive tension promotes symmetry-breaking.---}

\begin{figure}[ht!]
    \centering
    \includegraphics[width=8.6cm]{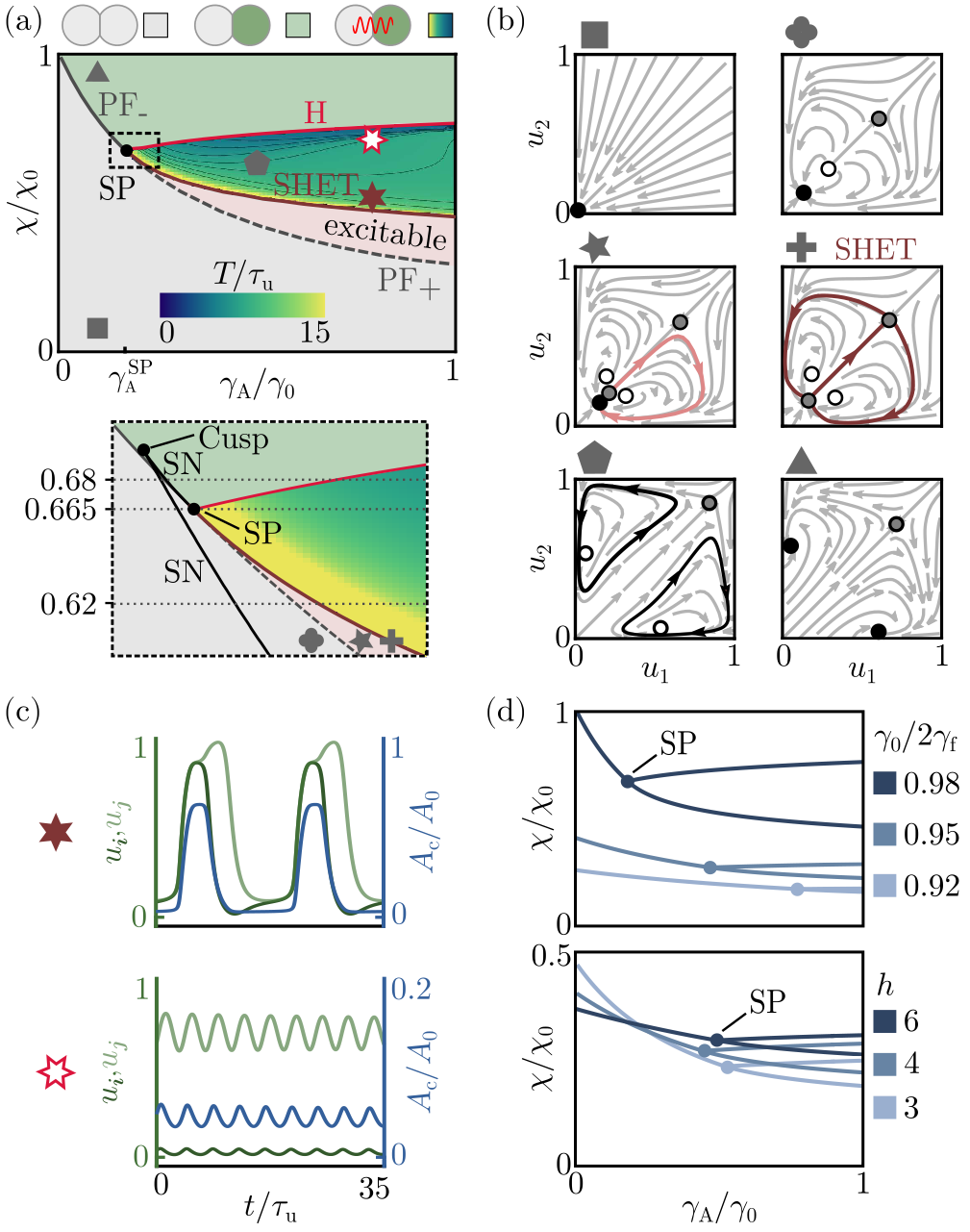}
    \caption{Excitability and oscillations of adaptive droplet pairs. (a) 
    Tension adaptation increases the parameter regime (green) of symmetry-breaking due to mutual inhibition (PF$_-$: supercritical pitchfork, PF$_+$: subcritical pitchfork), and leads to self-sustained oscillations of states and droplet shapes (color gradient and contour lines: oscillation period $T$). Saddle-heteroclinic (SHET) and Hopf (H) bifurcation lines bound the oscillatory regime, originating from a saddle-node pitchfork codimension-2 point (SP). Enlarged view of the SP point shows saddle-node (SN) and cusp bifurcations that preserve the stable attractors (reference susceptibility: critical value without adaptive tension $\chi_0=\left.\chi_{\rm{PF}}\right|_{\gamma_{\rm{A}}=0}$). (b) Phase portraits for parameter values marked in (a) (filled black circles: stable steady states, filled gray circles: saddles, open circles: unstable steady states, rose line: trajectory in the excitable regime, red lines: heteroclinics, black lines: limit cycles). (c) Oscillation amplitudes decrease with waveforms changing from relaxation-like (near SHET) to sinusoidal (near Hopf) for increasing $\chi$. (d) Top: The SP point shifts to the top left as $\gamma_0/2\gamma_{\rm{f}}$ increases. Bottom: Strong adaptive adhesion lowers symmetry-breaking and oscillation thresholds for small Hill coefficients.
    Parameter values given in \cite{sm}, Tab.~\ref{tab:parameter_values_figures}. For more details see companion paper~\cite[Fig.~5]{PREJoint}
    }
    \label{fig:symmetric_doublet_results}
\end{figure}

When droplets or cells exchange \emph{state-dependent} signals $\phi =\phi(u_1,\dots,u_N)$, the additional coupling between their internal states gives rise to new regimes. For example, strong \emph{mutually inhibitory} interactions generically lead to spontaneous symmetry-breaking, whereby initially small differences between interacting units diverge to low- and high-value steady states \cite{pisarchik2022multistability}, a mechanism relevant for the patterning of different cell types \cite{sjoqvist2019say}. 
To investigate how signal-induced adhesion affects symmetry-breaking, we performed numerical continuation of Eqs.~\eqref{eq:contact_area}--\eqref{eq:signal} (\cite{sm}, Sec.~\ref{sec:numerical_methods}) for droplet pairs with $\phi = \phi_0(1-u_j)$, an expression we derived from the microscopic dynamics of Delta-Notch signaling~\cite{PREJoint, sprinzak2021biophysics, khait2016quantitative}. Droplet pairs undergo symmetry-breaking of internal states above a line of supercritical pitchfork bifurcations (PF) in the state diagram spanned by the signal susceptibility $\chi$ and the adaptive adhesion coefficient $\gamma_{\rm{A}}$ [Fig.~\ref{fig:symmetric_doublet_results}(a) and (b) triangle]. Adaptive adhesion promotes symmetry-breaking by transiently expanding the contact area, which lowers the threshold susceptibility (Movie~1(f)). Adaptive contact dynamics indeed occur in developing zebrafish sensory cell pairs, which exchange mutually inhibitory signals across a transient contact surface to acquire distinct fates before detaching from each other~\cite{jacobo2019notch, erzberger2020mechanochemical}.

\paragraph*{Tunable self-sustained oscillations.---}For large adaptive adhesion coefficients, numerical continuation of Eqs.~\eqref{eq:contact_area}--\eqref{eq:signal} (\cite{sm}, Sec.~\ref{sec:numerical_methods}) reveals an oscillatory regime separating regions with symmetric and symmetry-broken steady states [Fig.~\ref{fig:symmetric_doublet_results}(a)], that originates from a saddle-node pitchfork bifurcation point (SP)---a codimension-2 bifurcation at which the PF line tangentially intersects with an SN bifurcation line (\cite{sm}, Fig.~\ref{fig:SPDetails} and \cite{PREJoint} for more details) \cite{blackbeard2012bursting}.
The oscillations are driven by the competition between shape dynamics and symmetry-breaking: the internal states increase the contact area according to Eq.~\ref{eq:tension_regulation}, thereby driving their own inhibition, leading to negative feedback. Simulations of Eqs.~\eqref{eq:contact_area}--\eqref{eq:signal} show characteristic dynamics near the Hopf (H) and saddle heteroclinic (SHET) bifurcation lines [Fig.~\ref{fig:symmetric_doublet_results}(c)] which delineate the oscillatory region. Near the transition between stable uniform and oscillatory states [Fig.~\ref{fig:symmetric_doublet_results}(a), SHET], droplet pairs become excitable: perturbations that cross the separatrices connecting the nearby saddle point to the unstable fixed points, trigger a large increase in contact area $A_{\rm{c}}$, followed by transient symmetry-breaking [Fig.~\ref{fig:symmetric_doublet_results}(b) star and Movie~1(b)]. At large $\chi$, the uniform stable fixed point and the saddle approach each other, which lowers the excitation threshold, until they collide [Fig.~\ref{fig:symmetric_doublet_results}(b) cross] and give rise to a pair of stable limit cycles [Fig.~\ref{fig:symmetric_doublet_results}(b) pentagon].
These cycles appear once transmitted signals become strong enough to induce symmetry-breaking entailing a decrease of adhesion and contact area, which in turn weakens signals below the symmetry-breaking threshold. Then the product of states increases again, and the cycle repeats.

The droplet oscillations exhibit a range of temporal profiles. Near the SHET line the droplet pair exhibits relaxation oscillations, which spend a large fraction of the cycle in the low-contact configuration, followed by spikes in the contact area $A_{\rm{c}}$ and rapid, transient symmetry-breaking [Fig.~\ref{fig:symmetric_doublet_results}(c)]. As $\chi$ approaches $\chi_{\rm{SHET}}$, the oscillation period diverges due to critical slowing near the ghost of the destroyed saddle point. Close to the Hopf bifurcation, oscillations approach harmonic waveforms, and limit cycles eventually contract into symmetry-broken fixed points [Fig.~\ref{fig:symmetric_doublet_results}(b) triangle].

The SP point's position and the size of associated regimes depend on the baseline tension ratio $\gamma_0/2\gamma_{\rm{f}}$, and the Hill coefficient $h$ [Eq.~\eqref{eq:u}]. Increasing $\gamma_0$ lowers the threshold adaptive tension for the onset of oscillations [Fig.~\ref{fig:symmetric_doublet_results}(d)], while for low $\gamma_0$ the adaptive adhesion can push the interface into an unstable regime where any area increase lowers the total surface energy~\cite{bormashenko2017physics, binysh2022active}. Close to $\gamma_{\rm{A}}/\gamma_0=1$, such instabilities may remain transient, and restabilize due to the decrease of adhesion upon symmetry-breaking of internal states, whereas at large $\gamma_{\rm{A}}/\gamma_0$, these effects are expected to lead to new phenomena.

We found shape bistabilities and symmetry-breaking for Hill coefficients $h\geq2$, and oscillations for $h\geq3$. Strongly nonlinear response functions are common in cellular regulatory feedbacks~\cite{boareto2015jagged, shi2018understanding, corson2017self}, and experimental evidence~\cite{dubrulle2015response} including our zebrafish embryo data~\cite[Fig.~\ref{fig:sbi_hill_analysis}(a)]{sm} suggests large Hill coefficients for Nodal signaling.
%Strongly nonlinear response functions are commonly used to model regulatory feedbacks in cells~\cite{boareto2015jagged, shi2018understanding, corson2017self}, and experimental evidence has been reported %for e.g. the Nodal pathways \cite{dubrulle2015response}, consistent with our analysis of zebrafish embryo data \cite[Fig.~\ref{fig:sbi_hill_analysis}(a)]{sm}.
Interestingly, strong adaptive adhesion achieves lower thresholds for smaller Hill coefficients, i.e. PF bifurcation lines for different Hill coefficients intersect in the feedback parameter space [Fig.~\ref{fig:symmetric_doublet_results}(d)], indicating a non-trivial interplay between the response nonlinearity and the geometry-dependent nonlinearity. 

Together, these results illustrate how shape adaptation serves as a mechanism of signal processing and generates complex temporal dynamics, such as excitability and self-sustained oscillations.

\paragraph*{Conclusions} 

Our analysis of adaptive, interacting droplets reveals the rich physics arising in signal-processing active materials. Coupling between droplet geometry and contact-dependent  signals drives shape bistability, robust symmetry-breaking, excitability, and oscillations of different waveforms---enabling a wide range of time-encoded outputs with minimal degrees of freedom. Dynamic signaling levels can indeed program distinct cell fates \cite{nandagopal2018dynamic, viswanathan2021desensitisation, sonnen2021signalling}, whereby shape-dependent feedback could self-organise multicellular structures~\cite{casani2022signaling, purvis2013encoding}. Applied to experimental observations from zebrafish embryos, our theory suggests that positive shape-adaptive feedback aids tissue boundary formation at the late blastoderm stage.

Our findings will enable the discovery of new collective phenomena in active signal-processing materials, where mechanical feedback drives spontaneous patterning and wave dynamics\cite{perez2023excitable, bailles2022mechanochemical, di2022waves}. Investigating fluctuation-induced effects in the excitable regimes for example could reveal how large area deviations trigger topological transitions governing global rheological properties \cite{corominas2021viscoelastic, guirao2017biomechanics}. Analysing these collective dynamics will uncover how geometry-dependent feedback produces novel modes of self-organisation in adaptive materials.
\newline

\paragraph*{Acknowledgements}
%------------------------------
We gratefully acknowledge insightful feedback from Florian Berger, Erwin Frey, Isabella Graf, Jeremy Gunawardena, Adrian Jacobo, Thomas Quail, Ulrich Schwarz, Alejandro Torres-Sánchez, Falko Ziebert and all members of the Erzberger Group. We acknowledge funding from EMBL and a Joachim-Herz Add-on fellowship for interdisciplinary life science for TD.
%------------------------------

% References
\bibliography{references.bib}

\begin{thebibliography}{81}%
\makeatletter
\providecommand \@ifxundefined [1]{%
 \@ifx{#1\undefined}
}%
\providecommand \@ifnum [1]{%
 \ifnum #1\expandafter \@firstoftwo
 \else \expandafter \@secondoftwo
 \fi
}%
\providecommand \@ifx [1]{%
 \ifx #1\expandafter \@firstoftwo
 \else \expandafter \@secondoftwo
 \fi
}%
\providecommand \natexlab [1]{#1}%
\providecommand \enquote  [1]{``#1''}%
\providecommand \bibnamefont  [1]{#1}%
\providecommand \bibfnamefont [1]{#1}%
\providecommand \citenamefont [1]{#1}%
\providecommand \href@noop [0]{\@secondoftwo}%
\providecommand \href [0]{\begingroup \@sanitize@url \@href}%
\providecommand \@href[1]{\@@startlink{#1}\@@href}%
\providecommand \@@href[1]{\endgroup#1\@@endlink}%
\providecommand \@sanitize@url [0]{\catcode `\\12\catcode `\$12\catcode `\&12\catcode `\#12\catcode `\^12\catcode `\_12\catcode `\%12\relax}%
\providecommand \@@startlink[1]{}%
\providecommand \@@endlink[0]{}%
\providecommand \url  [0]{\begingroup\@sanitize@url \@url }%
\providecommand \@url [1]{\endgroup\@href {#1}{\urlprefix }}%
\providecommand \urlprefix  [0]{URL }%
\providecommand \Eprint [0]{\href }%
\providecommand \doibase [0]{https://doi.org/}%
\providecommand \selectlanguage [0]{\@gobble}%
\providecommand \bibinfo  [0]{\@secondoftwo}%
\providecommand \bibfield  [0]{\@secondoftwo}%
\providecommand \translation [1]{[#1]}%
\providecommand \BibitemOpen [0]{}%
\providecommand \bibitemStop [0]{}%
\providecommand \bibitemNoStop [0]{.\EOS\space}%
\providecommand \EOS [0]{\spacefactor3000\relax}%
\providecommand \BibitemShut  [1]{\csname bibitem#1\endcsname}%
\let\auto@bib@innerbib\@empty
%</preamble>
\bibitem [{\citenamefont {Mietke}\ \emph {et~al.}(2019)\citenamefont {Mietke}, \citenamefont {J{\"u}licher},\ and\ \citenamefont {Sbalzarini}}]{mietke2019self}%
  \BibitemOpen
  \bibfield  {author} {\bibinfo {author} {\bibfnamefont {A.}~\bibnamefont {Mietke}}, \bibinfo {author} {\bibfnamefont {F.}~\bibnamefont {J{\"u}licher}},\ and\ \bibinfo {author} {\bibfnamefont {I.~F.}\ \bibnamefont {Sbalzarini}},\ }\href@noop {} {\bibfield  {journal} {\bibinfo  {journal} {Proceedings of the National Academy of Sciences}\ }\textbf {\bibinfo {volume} {116}},\ \bibinfo {pages} {29} (\bibinfo {year} {2019})}\BibitemShut {NoStop}%
\bibitem [{\citenamefont {Salbreux}\ and\ \citenamefont {J{\"u}licher}(2017)}]{salbreux2017mechanics}%
  \BibitemOpen
  \bibfield  {author} {\bibinfo {author} {\bibfnamefont {G.}~\bibnamefont {Salbreux}}\ and\ \bibinfo {author} {\bibfnamefont {F.}~\bibnamefont {J{\"u}licher}},\ }\href@noop {} {\bibfield  {journal} {\bibinfo  {journal} {Physical Review E}\ }\textbf {\bibinfo {volume} {96}},\ \bibinfo {pages} {032404} (\bibinfo {year} {2017})}\BibitemShut {NoStop}%
\bibitem [{\citenamefont {Marchetti}\ \emph {et~al.}(2013)\citenamefont {Marchetti}, \citenamefont {Joanny}, \citenamefont {Ramaswamy}, \citenamefont {Liverpool}, \citenamefont {Prost}, \citenamefont {Rao},\ and\ \citenamefont {Simha}}]{marchetti2013hydrodynamics}%
  \BibitemOpen
  \bibfield  {author} {\bibinfo {author} {\bibfnamefont {M.~C.}\ \bibnamefont {Marchetti}}, \bibinfo {author} {\bibfnamefont {J.-F.}\ \bibnamefont {Joanny}}, \bibinfo {author} {\bibfnamefont {S.}~\bibnamefont {Ramaswamy}}, \bibinfo {author} {\bibfnamefont {T.~B.}\ \bibnamefont {Liverpool}}, \bibinfo {author} {\bibfnamefont {J.}~\bibnamefont {Prost}}, \bibinfo {author} {\bibfnamefont {M.}~\bibnamefont {Rao}},\ and\ \bibinfo {author} {\bibfnamefont {R.~A.}\ \bibnamefont {Simha}},\ }\href@noop {} {\bibfield  {journal} {\bibinfo  {journal} {Reviews of modern physics}\ }\textbf {\bibinfo {volume} {85}},\ \bibinfo {pages} {1143} (\bibinfo {year} {2013})}\BibitemShut {NoStop}%
\bibitem [{\citenamefont {Shankar}\ \emph {et~al.}(2022)\citenamefont {Shankar}, \citenamefont {Souslov}, \citenamefont {Bowick}, \citenamefont {Marchetti},\ and\ \citenamefont {Vitelli}}]{shankar2022topological}%
  \BibitemOpen
  \bibfield  {author} {\bibinfo {author} {\bibfnamefont {S.}~\bibnamefont {Shankar}}, \bibinfo {author} {\bibfnamefont {A.}~\bibnamefont {Souslov}}, \bibinfo {author} {\bibfnamefont {M.~J.}\ \bibnamefont {Bowick}}, \bibinfo {author} {\bibfnamefont {M.~C.}\ \bibnamefont {Marchetti}},\ and\ \bibinfo {author} {\bibfnamefont {V.}~\bibnamefont {Vitelli}},\ }\href@noop {} {\bibfield  {journal} {\bibinfo  {journal} {Nature Reviews Physics}\ }\textbf {\bibinfo {volume} {4}},\ \bibinfo {pages} {380} (\bibinfo {year} {2022})}\BibitemShut {NoStop}%
\bibitem [{\citenamefont {Binysh}\ \emph {et~al.}(2022)\citenamefont {Binysh}, \citenamefont {Wilks},\ and\ \citenamefont {Souslov}}]{binysh2022active}%
  \BibitemOpen
  \bibfield  {author} {\bibinfo {author} {\bibfnamefont {J.}~\bibnamefont {Binysh}}, \bibinfo {author} {\bibfnamefont {T.~R.}\ \bibnamefont {Wilks}},\ and\ \bibinfo {author} {\bibfnamefont {A.}~\bibnamefont {Souslov}},\ }\href@noop {} {\bibfield  {journal} {\bibinfo  {journal} {Science advances}\ }\textbf {\bibinfo {volume} {8}},\ \bibinfo {pages} {eabk3079} (\bibinfo {year} {2022})}\BibitemShut {NoStop}%
\bibitem [{\citenamefont {Ray}\ \emph {et~al.}(2023)\citenamefont {Ray}, \citenamefont {Zhang},\ and\ \citenamefont {Dogic}}]{ray2023rectified}%
  \BibitemOpen
  \bibfield  {author} {\bibinfo {author} {\bibfnamefont {S.}~\bibnamefont {Ray}}, \bibinfo {author} {\bibfnamefont {J.}~\bibnamefont {Zhang}},\ and\ \bibinfo {author} {\bibfnamefont {Z.}~\bibnamefont {Dogic}},\ }\href@noop {} {\bibfield  {journal} {\bibinfo  {journal} {Physical Review Letters}\ }\textbf {\bibinfo {volume} {130}},\ \bibinfo {pages} {238301} (\bibinfo {year} {2023})}\BibitemShut {NoStop}%
\bibitem [{\citenamefont {Ara{\'u}jo}\ \emph {et~al.}(2023)\citenamefont {Ara{\'u}jo}, \citenamefont {Janssen}, \citenamefont {Barois}, \citenamefont {Boffetta}, \citenamefont {Cohen}, \citenamefont {Corbetta}, \citenamefont {Dauchot}, \citenamefont {Dijkstra}, \citenamefont {Durham}, \citenamefont {Dussutour} \emph {et~al.}}]{araujo2023steering}%
  \BibitemOpen
  \bibfield  {author} {\bibinfo {author} {\bibfnamefont {N.~A.}\ \bibnamefont {Ara{\'u}jo}}, \bibinfo {author} {\bibfnamefont {L.~M.}\ \bibnamefont {Janssen}}, \bibinfo {author} {\bibfnamefont {T.}~\bibnamefont {Barois}}, \bibinfo {author} {\bibfnamefont {G.}~\bibnamefont {Boffetta}}, \bibinfo {author} {\bibfnamefont {I.}~\bibnamefont {Cohen}}, \bibinfo {author} {\bibfnamefont {A.}~\bibnamefont {Corbetta}}, \bibinfo {author} {\bibfnamefont {O.}~\bibnamefont {Dauchot}}, \bibinfo {author} {\bibfnamefont {M.}~\bibnamefont {Dijkstra}}, \bibinfo {author} {\bibfnamefont {W.~M.}\ \bibnamefont {Durham}}, \bibinfo {author} {\bibfnamefont {A.}~\bibnamefont {Dussutour}}, \emph {et~al.},\ }\href@noop {} {\bibfield  {journal} {\bibinfo  {journal} {Soft matter}\ }\textbf {\bibinfo {volume} {19}},\ \bibinfo {pages} {1695} (\bibinfo {year} {2023})}\BibitemShut {NoStop}%
\bibitem [{\citenamefont {McEvoy}\ and\ \citenamefont {Correll}(2015)}]{mcevoy2015materials}%
  \BibitemOpen
  \bibfield  {author} {\bibinfo {author} {\bibfnamefont {M.~A.}\ \bibnamefont {McEvoy}}\ and\ \bibinfo {author} {\bibfnamefont {N.}~\bibnamefont {Correll}},\ }\href@noop {} {\bibfield  {journal} {\bibinfo  {journal} {Science}\ }\textbf {\bibinfo {volume} {347}},\ \bibinfo {pages} {1261689} (\bibinfo {year} {2015})}\BibitemShut {NoStop}%
\bibitem [{\citenamefont {Zhao}\ \emph {et~al.}(2023)\citenamefont {Zhao}, \citenamefont {Ko{\v{s}}mrlj},\ and\ \citenamefont {Datta}}]{zhao2023chemotactic}%
  \BibitemOpen
  \bibfield  {author} {\bibinfo {author} {\bibfnamefont {H.}~\bibnamefont {Zhao}}, \bibinfo {author} {\bibfnamefont {A.}~\bibnamefont {Ko{\v{s}}mrlj}},\ and\ \bibinfo {author} {\bibfnamefont {S.~S.}\ \bibnamefont {Datta}},\ }\href@noop {} {\bibfield  {journal} {\bibinfo  {journal} {arXiv preprint arXiv:2301.12345}\ } (\bibinfo {year} {2023})}\BibitemShut {NoStop}%
\bibitem [{\citenamefont {Ziepke}\ \emph {et~al.}(2022)\citenamefont {Ziepke}, \citenamefont {Maryshev}, \citenamefont {Aranson},\ and\ \citenamefont {Frey}}]{ziepke2022multi}%
  \BibitemOpen
  \bibfield  {author} {\bibinfo {author} {\bibfnamefont {A.}~\bibnamefont {Ziepke}}, \bibinfo {author} {\bibfnamefont {I.}~\bibnamefont {Maryshev}}, \bibinfo {author} {\bibfnamefont {I.~S.}\ \bibnamefont {Aranson}},\ and\ \bibinfo {author} {\bibfnamefont {E.}~\bibnamefont {Frey}},\ }\href@noop {} {\bibfield  {journal} {\bibinfo  {journal} {Nature communications}\ }\textbf {\bibinfo {volume} {13}},\ \bibinfo {pages} {6727} (\bibinfo {year} {2022})}\BibitemShut {NoStop}%
\bibitem [{\citenamefont {Gonzales}\ \emph {et~al.}(2023)\citenamefont {Gonzales}, \citenamefont {Suraritdechachai}, \citenamefont {Zechner},\ and\ \citenamefont {Tang}}]{gonzales2023bidirectional}%
  \BibitemOpen
  \bibfield  {author} {\bibinfo {author} {\bibfnamefont {D.~T.}\ \bibnamefont {Gonzales}}, \bibinfo {author} {\bibfnamefont {S.}~\bibnamefont {Suraritdechachai}}, \bibinfo {author} {\bibfnamefont {C.}~\bibnamefont {Zechner}},\ and\ \bibinfo {author} {\bibfnamefont {T.-Y.~D.}\ \bibnamefont {Tang}},\ }\href@noop {} {\bibfield  {journal} {\bibinfo  {journal} {ChemSystemsChem}\ }\textbf {\bibinfo {volume} {5}},\ \bibinfo {pages} {e202300029} (\bibinfo {year} {2023})}\BibitemShut {NoStop}%
\bibitem [{\citenamefont {Alberts}\ \emph {et~al.}(2022)\citenamefont {Alberts}, \citenamefont {Heald}, \citenamefont {Johnson}, \citenamefont {Morgan}, \citenamefont {Raff}, \citenamefont {Roberts},\ and\ \citenamefont {Walter}}]{alberts2022molecular}%
  \BibitemOpen
  \bibfield  {author} {\bibinfo {author} {\bibfnamefont {B.}~\bibnamefont {Alberts}}, \bibinfo {author} {\bibfnamefont {R.}~\bibnamefont {Heald}}, \bibinfo {author} {\bibfnamefont {A.}~\bibnamefont {Johnson}}, \bibinfo {author} {\bibfnamefont {D.}~\bibnamefont {Morgan}}, \bibinfo {author} {\bibfnamefont {M.}~\bibnamefont {Raff}}, \bibinfo {author} {\bibfnamefont {K.}~\bibnamefont {Roberts}},\ and\ \bibinfo {author} {\bibfnamefont {P.}~\bibnamefont {Walter}},\ }\href@noop {} {\emph {\bibinfo {title} {Molecular Biology of the Cell}}}\ (\bibinfo  {publisher} {WW Norton \& Company},\ \bibinfo {year} {2022})\BibitemShut {NoStop}%
\bibitem [{\citenamefont {Pisarchik}\ and\ \citenamefont {Hramov}(2022)}]{pisarchik2022multistability}%
  \BibitemOpen
  \bibfield  {author} {\bibinfo {author} {\bibfnamefont {A.~N.}\ \bibnamefont {Pisarchik}}\ and\ \bibinfo {author} {\bibfnamefont {A.~E.}\ \bibnamefont {Hramov}},\ }\href@noop {} {\emph {\bibinfo {title} {Multistability in Physical and Living Systems}}},\ Vol.~\bibinfo {volume} {2}\ (\bibinfo  {publisher} {Springer},\ \bibinfo {year} {2022})\ Chap.~\bibinfo {chapter} {3}\BibitemShut {NoStop}%
\bibitem [{\citenamefont {Sj{\"o}qvist}\ and\ \citenamefont {Andersson}(2019)}]{sjoqvist2019say}%
  \BibitemOpen
  \bibfield  {author} {\bibinfo {author} {\bibfnamefont {M.}~\bibnamefont {Sj{\"o}qvist}}\ and\ \bibinfo {author} {\bibfnamefont {E.~R.}\ \bibnamefont {Andersson}},\ }\href@noop {} {\bibfield  {journal} {\bibinfo  {journal} {Developmental biology}\ }\textbf {\bibinfo {volume} {447}},\ \bibinfo {pages} {58} (\bibinfo {year} {2019})}\BibitemShut {NoStop}%
\bibitem [{\citenamefont {Dullweber}\ and\ \citenamefont {Erzberger}(2023)}]{dullweber2023mechanochemical}%
  \BibitemOpen
  \bibfield  {author} {\bibinfo {author} {\bibfnamefont {T.}~\bibnamefont {Dullweber}}\ and\ \bibinfo {author} {\bibfnamefont {A.}~\bibnamefont {Erzberger}},\ }\href@noop {} {\bibfield  {journal} {\bibinfo  {journal} {Current Opinion in Systems Biology}\ ,\ \bibinfo {pages} {100445}} (\bibinfo {year} {2023})}\BibitemShut {NoStop}%
\bibitem [{\citenamefont {Sprinzak}\ and\ \citenamefont {Blacklow}(2021)}]{sprinzak2021biophysics}%
  \BibitemOpen
  \bibfield  {author} {\bibinfo {author} {\bibfnamefont {D.}~\bibnamefont {Sprinzak}}\ and\ \bibinfo {author} {\bibfnamefont {S.~C.}\ \bibnamefont {Blacklow}},\ }\href@noop {} {\bibfield  {journal} {\bibinfo  {journal} {Annual review of biophysics}\ }\textbf {\bibinfo {volume} {50}},\ \bibinfo {pages} {157} (\bibinfo {year} {2021})}\BibitemShut {NoStop}%
\bibitem [{\citenamefont {Erzberger}\ \emph {et~al.}(2020)\citenamefont {Erzberger}, \citenamefont {Jacobo}, \citenamefont {Dasgupta},\ and\ \citenamefont {Hudspeth}}]{erzberger2020mechanochemical}%
  \BibitemOpen
  \bibfield  {author} {\bibinfo {author} {\bibfnamefont {A.}~\bibnamefont {Erzberger}}, \bibinfo {author} {\bibfnamefont {A.}~\bibnamefont {Jacobo}}, \bibinfo {author} {\bibfnamefont {A.}~\bibnamefont {Dasgupta}},\ and\ \bibinfo {author} {\bibfnamefont {A.}~\bibnamefont {Hudspeth}},\ }\href@noop {} {\bibfield  {journal} {\bibinfo  {journal} {Nature physics}\ }\textbf {\bibinfo {volume} {16}},\ \bibinfo {pages} {949} (\bibinfo {year} {2020})}\BibitemShut {NoStop}%
\bibitem [{\citenamefont {Luo}\ \emph {et~al.}(2023)\citenamefont {Luo}, \citenamefont {Maheshwari}, \citenamefont {Danielescu}, \citenamefont {Li}, \citenamefont {Yang}, \citenamefont {Tao}, \citenamefont {Sun}, \citenamefont {Patel}, \citenamefont {Wang}, \citenamefont {Yang} \emph {et~al.}}]{luo2023autonomous}%
  \BibitemOpen
  \bibfield  {author} {\bibinfo {author} {\bibfnamefont {D.}~\bibnamefont {Luo}}, \bibinfo {author} {\bibfnamefont {A.}~\bibnamefont {Maheshwari}}, \bibinfo {author} {\bibfnamefont {A.}~\bibnamefont {Danielescu}}, \bibinfo {author} {\bibfnamefont {J.}~\bibnamefont {Li}}, \bibinfo {author} {\bibfnamefont {Y.}~\bibnamefont {Yang}}, \bibinfo {author} {\bibfnamefont {Y.}~\bibnamefont {Tao}}, \bibinfo {author} {\bibfnamefont {L.}~\bibnamefont {Sun}}, \bibinfo {author} {\bibfnamefont {D.~K.}\ \bibnamefont {Patel}}, \bibinfo {author} {\bibfnamefont {G.}~\bibnamefont {Wang}}, \bibinfo {author} {\bibfnamefont {S.}~\bibnamefont {Yang}}, \emph {et~al.},\ }\href@noop {} {\bibfield  {journal} {\bibinfo  {journal} {Nature}\ }\textbf {\bibinfo {volume} {614}},\ \bibinfo {pages} {463} (\bibinfo {year} {2023})}\BibitemShut {NoStop}%
\bibitem [{\citenamefont {Shah}\ \emph {et~al.}(2021)\citenamefont {Shah}, \citenamefont {Powers}, \citenamefont {Tilton}, \citenamefont {Kriegman}, \citenamefont {Bongard},\ and\ \citenamefont {Kramer-Bottiglio}}]{shah2021soft}%
  \BibitemOpen
  \bibfield  {author} {\bibinfo {author} {\bibfnamefont {D.~S.}\ \bibnamefont {Shah}}, \bibinfo {author} {\bibfnamefont {J.~P.}\ \bibnamefont {Powers}}, \bibinfo {author} {\bibfnamefont {L.~G.}\ \bibnamefont {Tilton}}, \bibinfo {author} {\bibfnamefont {S.}~\bibnamefont {Kriegman}}, \bibinfo {author} {\bibfnamefont {J.}~\bibnamefont {Bongard}},\ and\ \bibinfo {author} {\bibfnamefont {R.}~\bibnamefont {Kramer-Bottiglio}},\ }\href@noop {} {\bibfield  {journal} {\bibinfo  {journal} {Nature Machine Intelligence}\ }\textbf {\bibinfo {volume} {3}},\ \bibinfo {pages} {51} (\bibinfo {year} {2021})}\BibitemShut {NoStop}%
\bibitem [{\citenamefont {Venturini}\ \emph {et~al.}(2020)\citenamefont {Venturini}, \citenamefont {Pezzano}, \citenamefont {Catala~Castro}, \citenamefont {H{\"a}kkinen}, \citenamefont {Jim{\'e}nez-Delgado}, \citenamefont {Colomer-Rosell}, \citenamefont {Marro}, \citenamefont {Tolosa-Ramon}, \citenamefont {Paz-L{\'o}pez}, \citenamefont {Valverde} \emph {et~al.}}]{venturini2020nucleus}%
  \BibitemOpen
  \bibfield  {author} {\bibinfo {author} {\bibfnamefont {V.}~\bibnamefont {Venturini}}, \bibinfo {author} {\bibfnamefont {F.}~\bibnamefont {Pezzano}}, \bibinfo {author} {\bibfnamefont {F.}~\bibnamefont {Catala~Castro}}, \bibinfo {author} {\bibfnamefont {H.-M.}\ \bibnamefont {H{\"a}kkinen}}, \bibinfo {author} {\bibfnamefont {S.}~\bibnamefont {Jim{\'e}nez-Delgado}}, \bibinfo {author} {\bibfnamefont {M.}~\bibnamefont {Colomer-Rosell}}, \bibinfo {author} {\bibfnamefont {M.}~\bibnamefont {Marro}}, \bibinfo {author} {\bibfnamefont {Q.}~\bibnamefont {Tolosa-Ramon}}, \bibinfo {author} {\bibfnamefont {S.}~\bibnamefont {Paz-L{\'o}pez}}, \bibinfo {author} {\bibfnamefont {M.~A.}\ \bibnamefont {Valverde}}, \emph {et~al.},\ }\href@noop {} {\bibfield  {journal} {\bibinfo  {journal} {Science}\ }\textbf {\bibinfo {volume} {370}},\ \bibinfo {pages} {eaba2644} (\bibinfo {year} {2020})}\BibitemShut {NoStop}%
\bibitem [{\citenamefont {Bodor}\ \emph {et~al.}(2020)\citenamefont {Bodor}, \citenamefont {P{\"o}nisch}, \citenamefont {Endres},\ and\ \citenamefont {Paluch}}]{bodor2020cell}%
  \BibitemOpen
  \bibfield  {author} {\bibinfo {author} {\bibfnamefont {D.~L.}\ \bibnamefont {Bodor}}, \bibinfo {author} {\bibfnamefont {W.}~\bibnamefont {P{\"o}nisch}}, \bibinfo {author} {\bibfnamefont {R.~G.}\ \bibnamefont {Endres}},\ and\ \bibinfo {author} {\bibfnamefont {E.~K.}\ \bibnamefont {Paluch}},\ }\href@noop {} {\bibfield  {journal} {\bibinfo  {journal} {Developmental cell}\ }\textbf {\bibinfo {volume} {52}},\ \bibinfo {pages} {550} (\bibinfo {year} {2020})}\BibitemShut {NoStop}%
\bibitem [{\citenamefont {Terryn}\ \emph {et~al.}(2021)\citenamefont {Terryn}, \citenamefont {Langenbach}, \citenamefont {Roels}, \citenamefont {Brancart}, \citenamefont {Bakkali-Hassani}, \citenamefont {Poutrel}, \citenamefont {Georgopoulou}, \citenamefont {Thuruthel}, \citenamefont {Safaei}, \citenamefont {Ferrentino} \emph {et~al.}}]{terryn2021review}%
  \BibitemOpen
  \bibfield  {author} {\bibinfo {author} {\bibfnamefont {S.}~\bibnamefont {Terryn}}, \bibinfo {author} {\bibfnamefont {J.}~\bibnamefont {Langenbach}}, \bibinfo {author} {\bibfnamefont {E.}~\bibnamefont {Roels}}, \bibinfo {author} {\bibfnamefont {J.}~\bibnamefont {Brancart}}, \bibinfo {author} {\bibfnamefont {C.}~\bibnamefont {Bakkali-Hassani}}, \bibinfo {author} {\bibfnamefont {Q.-A.}\ \bibnamefont {Poutrel}}, \bibinfo {author} {\bibfnamefont {A.}~\bibnamefont {Georgopoulou}}, \bibinfo {author} {\bibfnamefont {T.~G.}\ \bibnamefont {Thuruthel}}, \bibinfo {author} {\bibfnamefont {A.}~\bibnamefont {Safaei}}, \bibinfo {author} {\bibfnamefont {P.}~\bibnamefont {Ferrentino}}, \emph {et~al.},\ }\href@noop {} {\bibfield  {journal} {\bibinfo  {journal} {Materials Today}\ }\textbf {\bibinfo {volume} {47}},\ \bibinfo {pages} {187} (\bibinfo {year} {2021})}\BibitemShut {NoStop}%
\bibitem [{\citenamefont {Ajeti}\ \emph {et~al.}(2019)\citenamefont {Ajeti}, \citenamefont {Tabatabai}, \citenamefont {Fleszar}, \citenamefont {Staddon}, \citenamefont {Seara}, \citenamefont {Suarez}, \citenamefont {Yousafzai}, \citenamefont {Bi}, \citenamefont {Kovar}, \citenamefont {Banerjee} \emph {et~al.}}]{ajeti2019wound}%
  \BibitemOpen
  \bibfield  {author} {\bibinfo {author} {\bibfnamefont {V.}~\bibnamefont {Ajeti}}, \bibinfo {author} {\bibfnamefont {A.~P.}\ \bibnamefont {Tabatabai}}, \bibinfo {author} {\bibfnamefont {A.~J.}\ \bibnamefont {Fleszar}}, \bibinfo {author} {\bibfnamefont {M.~F.}\ \bibnamefont {Staddon}}, \bibinfo {author} {\bibfnamefont {D.~S.}\ \bibnamefont {Seara}}, \bibinfo {author} {\bibfnamefont {C.}~\bibnamefont {Suarez}}, \bibinfo {author} {\bibfnamefont {M.~S.}\ \bibnamefont {Yousafzai}}, \bibinfo {author} {\bibfnamefont {D.}~\bibnamefont {Bi}}, \bibinfo {author} {\bibfnamefont {D.~R.}\ \bibnamefont {Kovar}}, \bibinfo {author} {\bibfnamefont {S.}~\bibnamefont {Banerjee}}, \emph {et~al.},\ }\href@noop {} {\bibfield  {journal} {\bibinfo  {journal} {Nature physics}\ }\textbf {\bibinfo {volume} {15}},\ \bibinfo {pages} {696} (\bibinfo {year} {2019})}\BibitemShut {NoStop}%
\bibitem [{\citenamefont {Sun}\ \emph {et~al.}(2023)\citenamefont {Sun}, \citenamefont {Zhou}, \citenamefont {Ma}, \citenamefont {Li}, \citenamefont {Gro{\ss}}, \citenamefont {Chen},\ and\ \citenamefont {Zhao}}]{sun2023mean}%
  \BibitemOpen
  \bibfield  {author} {\bibinfo {author} {\bibfnamefont {G.}~\bibnamefont {Sun}}, \bibinfo {author} {\bibfnamefont {R.}~\bibnamefont {Zhou}}, \bibinfo {author} {\bibfnamefont {Z.}~\bibnamefont {Ma}}, \bibinfo {author} {\bibfnamefont {Y.}~\bibnamefont {Li}}, \bibinfo {author} {\bibfnamefont {R.}~\bibnamefont {Gro{\ss}}}, \bibinfo {author} {\bibfnamefont {Z.}~\bibnamefont {Chen}},\ and\ \bibinfo {author} {\bibfnamefont {S.}~\bibnamefont {Zhao}},\ }\href@noop {} {\bibfield  {journal} {\bibinfo  {journal} {Nature Communications}\ }\textbf {\bibinfo {volume} {14}},\ \bibinfo {pages} {3476} (\bibinfo {year} {2023})}\BibitemShut {NoStop}%
\bibitem [{\citenamefont {Gov}(2018)}]{gov2018guided}%
  \BibitemOpen
  \bibfield  {author} {\bibinfo {author} {\bibfnamefont {N.}~\bibnamefont {Gov}},\ }\href@noop {} {\bibfield  {journal} {\bibinfo  {journal} {Philosophical Transactions of the Royal Society B: Biological Sciences}\ }\textbf {\bibinfo {volume} {373}},\ \bibinfo {pages} {20170115} (\bibinfo {year} {2018})}\BibitemShut {NoStop}%
\bibitem [{\citenamefont {Paluch}\ and\ \citenamefont {Heisenberg}(2009)}]{paluch2009biology}%
  \BibitemOpen
  \bibfield  {author} {\bibinfo {author} {\bibfnamefont {E.}~\bibnamefont {Paluch}}\ and\ \bibinfo {author} {\bibfnamefont {C.-P.}\ \bibnamefont {Heisenberg}},\ }\href@noop {} {\bibfield  {journal} {\bibinfo  {journal} {Current Biology}\ }\textbf {\bibinfo {volume} {19}},\ \bibinfo {pages} {R790} (\bibinfo {year} {2009})}\BibitemShut {NoStop}%
\bibitem [{\citenamefont {Linding}\ and\ \citenamefont {Klipp}(2021)}]{linding2021shapes}%
  \BibitemOpen
  \bibfield  {author} {\bibinfo {author} {\bibfnamefont {R.}~\bibnamefont {Linding}}\ and\ \bibinfo {author} {\bibfnamefont {E.}~\bibnamefont {Klipp}},\ }\href@noop {} {\bibfield  {journal} {\bibinfo  {journal} {Current Opinion in Systems Biology}\ }\textbf {\bibinfo {volume} {27}},\ \bibinfo {pages} {100354} (\bibinfo {year} {2021})}\BibitemShut {NoStop}%
\bibitem [{\citenamefont {Prasad}\ and\ \citenamefont {Alizadeh}(2019)}]{prasad2019cell}%
  \BibitemOpen
  \bibfield  {author} {\bibinfo {author} {\bibfnamefont {A.}~\bibnamefont {Prasad}}\ and\ \bibinfo {author} {\bibfnamefont {E.}~\bibnamefont {Alizadeh}},\ }\href@noop {} {\bibfield  {journal} {\bibinfo  {journal} {Trends in biotechnology}\ }\textbf {\bibinfo {volume} {37}},\ \bibinfo {pages} {347} (\bibinfo {year} {2019})}\BibitemShut {NoStop}%
\bibitem [{\citenamefont {Dullweber}\ \emph {et~al.}(2024{\natexlab{a}})\citenamefont {Dullweber}, \citenamefont {Belousov},\ and\ \citenamefont {Erzberger}}]{PREJoint}%
  \BibitemOpen
  \bibfield  {author} {\bibinfo {author} {\bibfnamefont {T.}~\bibnamefont {Dullweber}}, \bibinfo {author} {\bibfnamefont {R.}~\bibnamefont {Belousov}},\ and\ \bibinfo {author} {\bibfnamefont {A.}~\bibnamefont {Erzberger}},\ }\href@noop {} {\bibfield  {journal} {\bibinfo  {journal} {arXiv preprint arXiv:2411.15165}\ } (\bibinfo {year} {2024}{\natexlab{a}})}\BibitemShut {NoStop}%
\bibitem [{\citenamefont {Dullweber}\ \emph {et~al.}(2024{\natexlab{b}})\citenamefont {Dullweber}, \citenamefont {Belousov},\ and\ \citenamefont {Erzberger}}]{sm}%
  \BibitemOpen
  \bibfield  {author} {\bibinfo {author} {\bibfnamefont {T.}~\bibnamefont {Dullweber}}, \bibinfo {author} {\bibfnamefont {R.}~\bibnamefont {Belousov}},\ and\ \bibinfo {author} {\bibfnamefont {A.}~\bibnamefont {Erzberger}},\ }\href {https://git.embl.de/dullwebe/dullweber2024} {\bibinfo {title} {{The Supplemental Material contains a derivation of Eq. (2), a summary of the relevant biological background, and details on the numerical and experimental methodologies employed.}}} (\bibinfo {year} {2024}{\natexlab{b}})\BibitemShut {NoStop}%
\bibitem [{\citenamefont {Brakke}(1992)}]{brakke1992surface}%
  \BibitemOpen
  \bibfield  {author} {\bibinfo {author} {\bibfnamefont {K.~A.}\ \bibnamefont {Brakke}},\ }\href@noop {} {\bibfield  {journal} {\bibinfo  {journal} {Experimental mathematics}\ }\textbf {\bibinfo {volume} {1}},\ \bibinfo {pages} {141} (\bibinfo {year} {1992})}\BibitemShut {NoStop}%
\bibitem [{\citenamefont {Herszterg}\ \emph {et~al.}(2023)\citenamefont {Herszterg}, \citenamefont {de~Gennes}, \citenamefont {Cicolini}, \citenamefont {Huang}, \citenamefont {Alexandre}, \citenamefont {Smith}, \citenamefont {Araujo}, \citenamefont {Vincent},\ and\ \citenamefont {Salbreux}}]{herszterg2023signalling}%
  \BibitemOpen
  \bibfield  {author} {\bibinfo {author} {\bibfnamefont {S.}~\bibnamefont {Herszterg}}, \bibinfo {author} {\bibfnamefont {M.}~\bibnamefont {de~Gennes}}, \bibinfo {author} {\bibfnamefont {S.}~\bibnamefont {Cicolini}}, \bibinfo {author} {\bibfnamefont {A.}~\bibnamefont {Huang}}, \bibinfo {author} {\bibfnamefont {C.}~\bibnamefont {Alexandre}}, \bibinfo {author} {\bibfnamefont {M.}~\bibnamefont {Smith}}, \bibinfo {author} {\bibfnamefont {H.}~\bibnamefont {Araujo}}, \bibinfo {author} {\bibfnamefont {J.-P.}\ \bibnamefont {Vincent}},\ and\ \bibinfo {author} {\bibfnamefont {G.}~\bibnamefont {Salbreux}},\ }\href@noop {} {\bibfield  {journal} {\bibinfo  {journal} {bioRxiv}\ ,\ \bibinfo {pages} {2023}} (\bibinfo {year} {2023})}\BibitemShut {NoStop}%
\bibitem [{\citenamefont {Corson}\ \emph {et~al.}(2017)\citenamefont {Corson}, \citenamefont {Couturier}, \citenamefont {Rouault}, \citenamefont {Mazouni},\ and\ \citenamefont {Schweisguth}}]{corson2017self}%
  \BibitemOpen
  \bibfield  {author} {\bibinfo {author} {\bibfnamefont {F.}~\bibnamefont {Corson}}, \bibinfo {author} {\bibfnamefont {L.}~\bibnamefont {Couturier}}, \bibinfo {author} {\bibfnamefont {H.}~\bibnamefont {Rouault}}, \bibinfo {author} {\bibfnamefont {K.}~\bibnamefont {Mazouni}},\ and\ \bibinfo {author} {\bibfnamefont {F.}~\bibnamefont {Schweisguth}},\ }\href@noop {} {\bibfield  {journal} {\bibinfo  {journal} {Science}\ }\textbf {\bibinfo {volume} {356}},\ \bibinfo {pages} {eaai7407} (\bibinfo {year} {2017})}\BibitemShut {NoStop}%
\bibitem [{\citenamefont {Sitarska}\ and\ \citenamefont {Diz-Mu{\~n}oz}(2020)}]{sitarska2020pay}%
  \BibitemOpen
  \bibfield  {author} {\bibinfo {author} {\bibfnamefont {E.}~\bibnamefont {Sitarska}}\ and\ \bibinfo {author} {\bibfnamefont {A.}~\bibnamefont {Diz-Mu{\~n}oz}},\ }\href@noop {} {\bibfield  {journal} {\bibinfo  {journal} {Current opinion in cell biology}\ }\textbf {\bibinfo {volume} {66}},\ \bibinfo {pages} {11} (\bibinfo {year} {2020})}\BibitemShut {NoStop}%
\bibitem [{\citenamefont {Chugh}\ \emph {et~al.}(2017)\citenamefont {Chugh}, \citenamefont {Clark}, \citenamefont {Smith}, \citenamefont {Cassani}, \citenamefont {Dierkes}, \citenamefont {Ragab}, \citenamefont {Roux}, \citenamefont {Charras}, \citenamefont {Salbreux},\ and\ \citenamefont {Paluch}}]{chugh2017actin}%
  \BibitemOpen
  \bibfield  {author} {\bibinfo {author} {\bibfnamefont {P.}~\bibnamefont {Chugh}}, \bibinfo {author} {\bibfnamefont {A.~G.}\ \bibnamefont {Clark}}, \bibinfo {author} {\bibfnamefont {M.~B.}\ \bibnamefont {Smith}}, \bibinfo {author} {\bibfnamefont {D.~A.}\ \bibnamefont {Cassani}}, \bibinfo {author} {\bibfnamefont {K.}~\bibnamefont {Dierkes}}, \bibinfo {author} {\bibfnamefont {A.}~\bibnamefont {Ragab}}, \bibinfo {author} {\bibfnamefont {P.~P.}\ \bibnamefont {Roux}}, \bibinfo {author} {\bibfnamefont {G.}~\bibnamefont {Charras}}, \bibinfo {author} {\bibfnamefont {G.}~\bibnamefont {Salbreux}},\ and\ \bibinfo {author} {\bibfnamefont {E.~K.}\ \bibnamefont {Paluch}},\ }\href@noop {} {\bibfield  {journal} {\bibinfo  {journal} {Nature cell biology}\ }\textbf {\bibinfo {volume} {19}},\ \bibinfo {pages} {689} (\bibinfo {year} {2017})}\BibitemShut {NoStop}%
\bibitem [{\citenamefont {Nambiar}\ \emph {et~al.}(2009)\citenamefont {Nambiar}, \citenamefont {McConnell},\ and\ \citenamefont {Tyska}}]{nambiar2009control}%
  \BibitemOpen
  \bibfield  {author} {\bibinfo {author} {\bibfnamefont {R.}~\bibnamefont {Nambiar}}, \bibinfo {author} {\bibfnamefont {R.~E.}\ \bibnamefont {McConnell}},\ and\ \bibinfo {author} {\bibfnamefont {M.~J.}\ \bibnamefont {Tyska}},\ }\href@noop {} {\bibfield  {journal} {\bibinfo  {journal} {Proceedings of the National Academy of Sciences}\ }\textbf {\bibinfo {volume} {106}},\ \bibinfo {pages} {11972} (\bibinfo {year} {2009})}\BibitemShut {NoStop}%
\bibitem [{\citenamefont {Schwarz}\ and\ \citenamefont {Safran}(2013)}]{schwarz2013physics}%
  \BibitemOpen
  \bibfield  {author} {\bibinfo {author} {\bibfnamefont {U.~S.}\ \bibnamefont {Schwarz}}\ and\ \bibinfo {author} {\bibfnamefont {S.~A.}\ \bibnamefont {Safran}},\ }\href@noop {} {\bibfield  {journal} {\bibinfo  {journal} {Reviews of Modern Physics}\ }\textbf {\bibinfo {volume} {85}},\ \bibinfo {pages} {1327} (\bibinfo {year} {2013})}\BibitemShut {NoStop}%
\bibitem [{\citenamefont {Ma{\^\i}tre}\ \emph {et~al.}(2012)\citenamefont {Ma{\^\i}tre}, \citenamefont {Berthoumieux}, \citenamefont {Krens}, \citenamefont {Salbreux}, \citenamefont {J{\"u}licher}, \citenamefont {Paluch},\ and\ \citenamefont {Heisenberg}}]{maitre2012adhesion}%
  \BibitemOpen
  \bibfield  {author} {\bibinfo {author} {\bibfnamefont {J.-L.}\ \bibnamefont {Ma{\^\i}tre}}, \bibinfo {author} {\bibfnamefont {H.}~\bibnamefont {Berthoumieux}}, \bibinfo {author} {\bibfnamefont {S.~F.~G.}\ \bibnamefont {Krens}}, \bibinfo {author} {\bibfnamefont {G.}~\bibnamefont {Salbreux}}, \bibinfo {author} {\bibfnamefont {F.}~\bibnamefont {J{\"u}licher}}, \bibinfo {author} {\bibfnamefont {E.}~\bibnamefont {Paluch}},\ and\ \bibinfo {author} {\bibfnamefont {C.-P.}\ \bibnamefont {Heisenberg}},\ }\href@noop {} {\bibfield  {journal} {\bibinfo  {journal} {science}\ }\textbf {\bibinfo {volume} {338}},\ \bibinfo {pages} {253} (\bibinfo {year} {2012})}\BibitemShut {NoStop}%
\bibitem [{\citenamefont {Collier}\ \emph {et~al.}(1996)\citenamefont {Collier}, \citenamefont {Monk}, \citenamefont {Maini},\ and\ \citenamefont {Lewis}}]{collier1996pattern}%
  \BibitemOpen
  \bibfield  {author} {\bibinfo {author} {\bibfnamefont {J.~R.}\ \bibnamefont {Collier}}, \bibinfo {author} {\bibfnamefont {N.~A.}\ \bibnamefont {Monk}}, \bibinfo {author} {\bibfnamefont {P.~K.}\ \bibnamefont {Maini}},\ and\ \bibinfo {author} {\bibfnamefont {J.~H.}\ \bibnamefont {Lewis}},\ }\href@noop {} {\bibfield  {journal} {\bibinfo  {journal} {Journal of Theoretical Biology}\ }\textbf {\bibinfo {volume} {183}},\ \bibinfo {pages} {429} (\bibinfo {year} {1996})}\BibitemShut {NoStop}%
\bibitem [{\citenamefont {Binshtok}\ and\ \citenamefont {Sprinzak}(2018)}]{binshtok2018modeling}%
  \BibitemOpen
  \bibfield  {author} {\bibinfo {author} {\bibfnamefont {U.}~\bibnamefont {Binshtok}}\ and\ \bibinfo {author} {\bibfnamefont {D.}~\bibnamefont {Sprinzak}},\ }\href@noop {} {\bibfield  {journal} {\bibinfo  {journal} {Molecular Mechanisms of Notch Signaling}\ ,\ \bibinfo {pages} {79}} (\bibinfo {year} {2018})}\BibitemShut {NoStop}%
\bibitem [{\citenamefont {Khait}\ \emph {et~al.}(2016)\citenamefont {Khait}, \citenamefont {Orsher}, \citenamefont {Golan}, \citenamefont {Binshtok}, \citenamefont {Gordon-Bar}, \citenamefont {Amir-Zilberstein},\ and\ \citenamefont {Sprinzak}}]{khait2016quantitative}%
  \BibitemOpen
  \bibfield  {author} {\bibinfo {author} {\bibfnamefont {I.}~\bibnamefont {Khait}}, \bibinfo {author} {\bibfnamefont {Y.}~\bibnamefont {Orsher}}, \bibinfo {author} {\bibfnamefont {O.}~\bibnamefont {Golan}}, \bibinfo {author} {\bibfnamefont {U.}~\bibnamefont {Binshtok}}, \bibinfo {author} {\bibfnamefont {N.}~\bibnamefont {Gordon-Bar}}, \bibinfo {author} {\bibfnamefont {L.}~\bibnamefont {Amir-Zilberstein}},\ and\ \bibinfo {author} {\bibfnamefont {D.}~\bibnamefont {Sprinzak}},\ }\href@noop {} {\bibfield  {journal} {\bibinfo  {journal} {Cell reports}\ }\textbf {\bibinfo {volume} {14}},\ \bibinfo {pages} {225} (\bibinfo {year} {2016})}\BibitemShut {NoStop}%
\bibitem [{\citenamefont {Shaya}\ \emph {et~al.}(2017)\citenamefont {Shaya}, \citenamefont {Binshtok}, \citenamefont {Hersch}, \citenamefont {Rivkin}, \citenamefont {Weinreb}, \citenamefont {Amir-Zilberstein}, \citenamefont {Khamaisi}, \citenamefont {Oppenheim}, \citenamefont {Desai}, \citenamefont {Goodyear} \emph {et~al.}}]{shaya2017cell}%
  \BibitemOpen
  \bibfield  {author} {\bibinfo {author} {\bibfnamefont {O.}~\bibnamefont {Shaya}}, \bibinfo {author} {\bibfnamefont {U.}~\bibnamefont {Binshtok}}, \bibinfo {author} {\bibfnamefont {M.}~\bibnamefont {Hersch}}, \bibinfo {author} {\bibfnamefont {D.}~\bibnamefont {Rivkin}}, \bibinfo {author} {\bibfnamefont {S.}~\bibnamefont {Weinreb}}, \bibinfo {author} {\bibfnamefont {L.}~\bibnamefont {Amir-Zilberstein}}, \bibinfo {author} {\bibfnamefont {B.}~\bibnamefont {Khamaisi}}, \bibinfo {author} {\bibfnamefont {O.}~\bibnamefont {Oppenheim}}, \bibinfo {author} {\bibfnamefont {R.~A.}\ \bibnamefont {Desai}}, \bibinfo {author} {\bibfnamefont {R.~J.}\ \bibnamefont {Goodyear}}, \emph {et~al.},\ }\href@noop {} {\bibfield  {journal} {\bibinfo  {journal} {Developmental cell}\ }\textbf {\bibinfo {volume} {40}},\ \bibinfo {pages} {505} (\bibinfo {year} {2017})}\BibitemShut {NoStop}%
\bibitem [{\citenamefont {Wyatt}\ \emph {et~al.}(2016)\citenamefont {Wyatt}, \citenamefont {Baum},\ and\ \citenamefont {Charras}}]{wyatt2016question}%
  \BibitemOpen
  \bibfield  {author} {\bibinfo {author} {\bibfnamefont {T.}~\bibnamefont {Wyatt}}, \bibinfo {author} {\bibfnamefont {B.}~\bibnamefont {Baum}},\ and\ \bibinfo {author} {\bibfnamefont {G.}~\bibnamefont {Charras}},\ }\href@noop {} {\bibfield  {journal} {\bibinfo  {journal} {Current opinion in cell biology}\ }\textbf {\bibinfo {volume} {38}},\ \bibinfo {pages} {68} (\bibinfo {year} {2016})}\BibitemShut {NoStop}%
\bibitem [{\citenamefont {Tran-Son-Tay}\ \emph {et~al.}(1991)\citenamefont {Tran-Son-Tay}, \citenamefont {Needham}, \citenamefont {Yeung},\ and\ \citenamefont {Hochmuth}}]{tran1991time}%
  \BibitemOpen
  \bibfield  {author} {\bibinfo {author} {\bibfnamefont {R.}~\bibnamefont {Tran-Son-Tay}}, \bibinfo {author} {\bibfnamefont {D.}~\bibnamefont {Needham}}, \bibinfo {author} {\bibfnamefont {A.}~\bibnamefont {Yeung}},\ and\ \bibinfo {author} {\bibfnamefont {R.}~\bibnamefont {Hochmuth}},\ }\href@noop {} {\bibfield  {journal} {\bibinfo  {journal} {Biophysical journal}\ }\textbf {\bibinfo {volume} {60}},\ \bibinfo {pages} {856} (\bibinfo {year} {1991})}\BibitemShut {NoStop}%
\bibitem [{\citenamefont {Buccitelli}\ and\ \citenamefont {Selbach}(2020)}]{buccitelli2020mrnas}%
  \BibitemOpen
  \bibfield  {author} {\bibinfo {author} {\bibfnamefont {C.}~\bibnamefont {Buccitelli}}\ and\ \bibinfo {author} {\bibfnamefont {M.}~\bibnamefont {Selbach}},\ }\href@noop {} {\bibfield  {journal} {\bibinfo  {journal} {Nature Reviews Genetics}\ }\textbf {\bibinfo {volume} {21}},\ \bibinfo {pages} {630} (\bibinfo {year} {2020})}\BibitemShut {NoStop}%
\bibitem [{\citenamefont {Shamir}\ \emph {et~al.}(2016)\citenamefont {Shamir}, \citenamefont {Bar-On}, \citenamefont {Phillips},\ and\ \citenamefont {Milo}}]{SHAMIR20161302}%
  \BibitemOpen
  \bibfield  {author} {\bibinfo {author} {\bibfnamefont {M.}~\bibnamefont {Shamir}}, \bibinfo {author} {\bibfnamefont {Y.}~\bibnamefont {Bar-On}}, \bibinfo {author} {\bibfnamefont {R.}~\bibnamefont {Phillips}},\ and\ \bibinfo {author} {\bibfnamefont {R.}~\bibnamefont {Milo}},\ }\href {https://doi.org/https://doi.org/10.1016/j.cell.2016.02.058} {\bibfield  {journal} {\bibinfo  {journal} {Cell}\ }\textbf {\bibinfo {volume} {164}},\ \bibinfo {pages} {1302} (\bibinfo {year} {2016})}\BibitemShut {NoStop}%
\bibitem [{\citenamefont {Tejero-Cantero}\ \emph {et~al.}(2020)\citenamefont {Tejero-Cantero}, \citenamefont {Boelts}, \citenamefont {Deistler}, \citenamefont {Lueckmann}, \citenamefont {Durkan}, \citenamefont {Gonçalves}, \citenamefont {Greenberg},\ and\ \citenamefont {Macke}}]{Tejero-Cantero2020}%
  \BibitemOpen
  \bibfield  {author} {\bibinfo {author} {\bibfnamefont {A.}~\bibnamefont {Tejero-Cantero}}, \bibinfo {author} {\bibfnamefont {J.}~\bibnamefont {Boelts}}, \bibinfo {author} {\bibfnamefont {M.}~\bibnamefont {Deistler}}, \bibinfo {author} {\bibfnamefont {J.-M.}\ \bibnamefont {Lueckmann}}, \bibinfo {author} {\bibfnamefont {C.}~\bibnamefont {Durkan}}, \bibinfo {author} {\bibfnamefont {P.~J.}\ \bibnamefont {Gonçalves}}, \bibinfo {author} {\bibfnamefont {D.~S.}\ \bibnamefont {Greenberg}},\ and\ \bibinfo {author} {\bibfnamefont {J.~H.}\ \bibnamefont {Macke}},\ }\href {https://doi.org/10.21105/joss.02505} {\bibfield  {journal} {\bibinfo  {journal} {Journal of Open Source Software}\ }\textbf {\bibinfo {volume} {5}},\ \bibinfo {pages} {2505} (\bibinfo {year} {2020})}\BibitemShut {NoStop}%
\bibitem [{\citenamefont {Hill}(2022)}]{hill2022establishment}%
  \BibitemOpen
  \bibfield  {author} {\bibinfo {author} {\bibfnamefont {C.~S.}\ \bibnamefont {Hill}},\ }\href@noop {} {\bibfield  {journal} {\bibinfo  {journal} {Current topics in developmental biology}\ }\textbf {\bibinfo {volume} {149}},\ \bibinfo {pages} {311} (\bibinfo {year} {2022})}\BibitemShut {NoStop}%
\bibitem [{\citenamefont {M{\"u}ller}\ \emph {et~al.}(2012)\citenamefont {M{\"u}ller}, \citenamefont {Rogers}, \citenamefont {Jordan}, \citenamefont {Lee}, \citenamefont {Robson}, \citenamefont {Ramanathan},\ and\ \citenamefont {Schier}}]{muller2012differential}%
  \BibitemOpen
  \bibfield  {author} {\bibinfo {author} {\bibfnamefont {P.}~\bibnamefont {M{\"u}ller}}, \bibinfo {author} {\bibfnamefont {K.~W.}\ \bibnamefont {Rogers}}, \bibinfo {author} {\bibfnamefont {B.~M.}\ \bibnamefont {Jordan}}, \bibinfo {author} {\bibfnamefont {J.~S.}\ \bibnamefont {Lee}}, \bibinfo {author} {\bibfnamefont {D.}~\bibnamefont {Robson}}, \bibinfo {author} {\bibfnamefont {S.}~\bibnamefont {Ramanathan}},\ and\ \bibinfo {author} {\bibfnamefont {A.~F.}\ \bibnamefont {Schier}},\ }\href@noop {} {\bibfield  {journal} {\bibinfo  {journal} {Science}\ }\textbf {\bibinfo {volume} {336}},\ \bibinfo {pages} {721} (\bibinfo {year} {2012})}\BibitemShut {NoStop}%
\bibitem [{\citenamefont {Petridou}\ \emph {et~al.}(2021)\citenamefont {Petridou}, \citenamefont {Corominas-Murtra}, \citenamefont {Heisenberg},\ and\ \citenamefont {Hannezo}}]{petridou2021rigidity}%
  \BibitemOpen
  \bibfield  {author} {\bibinfo {author} {\bibfnamefont {N.~I.}\ \bibnamefont {Petridou}}, \bibinfo {author} {\bibfnamefont {B.}~\bibnamefont {Corominas-Murtra}}, \bibinfo {author} {\bibfnamefont {C.-P.}\ \bibnamefont {Heisenberg}},\ and\ \bibinfo {author} {\bibfnamefont {E.}~\bibnamefont {Hannezo}},\ }\href@noop {} {\bibfield  {journal} {\bibinfo  {journal} {Cell}\ }\textbf {\bibinfo {volume} {184}},\ \bibinfo {pages} {1914} (\bibinfo {year} {2021})}\BibitemShut {NoStop}%
\bibitem [{\citenamefont {Barone}\ \emph {et~al.}(2017)\citenamefont {Barone}, \citenamefont {Lang}, \citenamefont {Krens}, \citenamefont {Pradhan}, \citenamefont {Shamipour}, \citenamefont {Sako}, \citenamefont {Sikora}, \citenamefont {Guet},\ and\ \citenamefont {Heisenberg}}]{barone2017effective}%
  \BibitemOpen
  \bibfield  {author} {\bibinfo {author} {\bibfnamefont {V.}~\bibnamefont {Barone}}, \bibinfo {author} {\bibfnamefont {M.}~\bibnamefont {Lang}}, \bibinfo {author} {\bibfnamefont {S.~G.}\ \bibnamefont {Krens}}, \bibinfo {author} {\bibfnamefont {S.~J.}\ \bibnamefont {Pradhan}}, \bibinfo {author} {\bibfnamefont {S.}~\bibnamefont {Shamipour}}, \bibinfo {author} {\bibfnamefont {K.}~\bibnamefont {Sako}}, \bibinfo {author} {\bibfnamefont {M.}~\bibnamefont {Sikora}}, \bibinfo {author} {\bibfnamefont {C.~C.}\ \bibnamefont {Guet}},\ and\ \bibinfo {author} {\bibfnamefont {C.-P.}\ \bibnamefont {Heisenberg}},\ }\href@noop {} {\bibfield  {journal} {\bibinfo  {journal} {Developmental cell}\ }\textbf {\bibinfo {volume} {43}},\ \bibinfo {pages} {198} (\bibinfo {year} {2017})}\BibitemShut {NoStop}%
\bibitem [{\citenamefont {Drenckhan}\ and\ \citenamefont {Hutzler}(2015)}]{drenckhan2015structure}%
  \BibitemOpen
  \bibfield  {author} {\bibinfo {author} {\bibfnamefont {W.}~\bibnamefont {Drenckhan}}\ and\ \bibinfo {author} {\bibfnamefont {S.}~\bibnamefont {Hutzler}},\ }\href@noop {} {\bibfield  {journal} {\bibinfo  {journal} {Advances in colloid and interface science}\ }\textbf {\bibinfo {volume} {224}},\ \bibinfo {pages} {1} (\bibinfo {year} {2015})}\BibitemShut {NoStop}%
\bibitem [{\citenamefont {Rogers}\ and\ \citenamefont {Schier}(2011)}]{rogers2011morphogen}%
  \BibitemOpen
  \bibfield  {author} {\bibinfo {author} {\bibfnamefont {K.~W.}\ \bibnamefont {Rogers}}\ and\ \bibinfo {author} {\bibfnamefont {A.~F.}\ \bibnamefont {Schier}},\ }\href@noop {} {\bibfield  {journal} {\bibinfo  {journal} {Annual review of cell and developmental biology}\ }\textbf {\bibinfo {volume} {27}},\ \bibinfo {pages} {377} (\bibinfo {year} {2011})}\BibitemShut {NoStop}%
\bibitem [{\citenamefont {Gregor}\ \emph {et~al.}(2007)\citenamefont {Gregor}, \citenamefont {Tank}, \citenamefont {Wieschaus},\ and\ \citenamefont {Bialek}}]{gregor2007probing}%
  \BibitemOpen
  \bibfield  {author} {\bibinfo {author} {\bibfnamefont {T.}~\bibnamefont {Gregor}}, \bibinfo {author} {\bibfnamefont {D.~W.}\ \bibnamefont {Tank}}, \bibinfo {author} {\bibfnamefont {E.~F.}\ \bibnamefont {Wieschaus}},\ and\ \bibinfo {author} {\bibfnamefont {W.}~\bibnamefont {Bialek}},\ }\href@noop {} {\bibfield  {journal} {\bibinfo  {journal} {Cell}\ }\textbf {\bibinfo {volume} {130}},\ \bibinfo {pages} {153} (\bibinfo {year} {2007})}\BibitemShut {NoStop}%
\bibitem [{\citenamefont {Chandrasekhar}(1943)}]{chandrasekhar1943stochastic}%
  \BibitemOpen
  \bibfield  {author} {\bibinfo {author} {\bibfnamefont {S.}~\bibnamefont {Chandrasekhar}},\ }\href@noop {} {\bibfield  {journal} {\bibinfo  {journal} {Reviews of modern physics}\ }\textbf {\bibinfo {volume} {15}},\ \bibinfo {pages} {1} (\bibinfo {year} {1943})}\BibitemShut {NoStop}%
\bibitem [{\citenamefont {Witzel}\ \emph {et~al.}(2006)\citenamefont {Witzel}, \citenamefont {Zimyanin}, \citenamefont {Carreira-Barbosa}, \citenamefont {Tada},\ and\ \citenamefont {Heisenberg}}]{witzel2006wnt11}%
  \BibitemOpen
  \bibfield  {author} {\bibinfo {author} {\bibfnamefont {S.}~\bibnamefont {Witzel}}, \bibinfo {author} {\bibfnamefont {V.}~\bibnamefont {Zimyanin}}, \bibinfo {author} {\bibfnamefont {F.}~\bibnamefont {Carreira-Barbosa}}, \bibinfo {author} {\bibfnamefont {M.}~\bibnamefont {Tada}},\ and\ \bibinfo {author} {\bibfnamefont {C.-P.}\ \bibnamefont {Heisenberg}},\ }\href@noop {} {\bibfield  {journal} {\bibinfo  {journal} {The Journal of cell biology}\ }\textbf {\bibinfo {volume} {175}},\ \bibinfo {pages} {791} (\bibinfo {year} {2006})}\BibitemShut {NoStop}%
\bibitem [{\citenamefont {Ulrich}\ \emph {et~al.}(2005)\citenamefont {Ulrich}, \citenamefont {Krieg}, \citenamefont {Sch{\"o}tz}, \citenamefont {Link}, \citenamefont {Castanon}, \citenamefont {Schnabel}, \citenamefont {Taubenberger}, \citenamefont {Mueller}, \citenamefont {Puech},\ and\ \citenamefont {Heisenberg}}]{ulrich2005wnt11}%
  \BibitemOpen
  \bibfield  {author} {\bibinfo {author} {\bibfnamefont {F.}~\bibnamefont {Ulrich}}, \bibinfo {author} {\bibfnamefont {M.}~\bibnamefont {Krieg}}, \bibinfo {author} {\bibfnamefont {E.-M.}\ \bibnamefont {Sch{\"o}tz}}, \bibinfo {author} {\bibfnamefont {V.}~\bibnamefont {Link}}, \bibinfo {author} {\bibfnamefont {I.}~\bibnamefont {Castanon}}, \bibinfo {author} {\bibfnamefont {V.}~\bibnamefont {Schnabel}}, \bibinfo {author} {\bibfnamefont {A.}~\bibnamefont {Taubenberger}}, \bibinfo {author} {\bibfnamefont {D.}~\bibnamefont {Mueller}}, \bibinfo {author} {\bibfnamefont {P.-H.}\ \bibnamefont {Puech}},\ and\ \bibinfo {author} {\bibfnamefont {C.-P.}\ \bibnamefont {Heisenberg}},\ }\href@noop {} {\bibfield  {journal} {\bibinfo  {journal} {Developmental cell}\ }\textbf {\bibinfo {volume} {9}},\ \bibinfo {pages} {555} (\bibinfo {year} {2005})}\BibitemShut {NoStop}%
\bibitem [{\citenamefont {van Boxtel}\ \emph {et~al.}(2015)\citenamefont {van Boxtel}, \citenamefont {Chesebro}, \citenamefont {Heliot}, \citenamefont {Ramel}, \citenamefont {Stone},\ and\ \citenamefont {Hill}}]{van2015temporal}%
  \BibitemOpen
  \bibfield  {author} {\bibinfo {author} {\bibfnamefont {A.~L.}\ \bibnamefont {van Boxtel}}, \bibinfo {author} {\bibfnamefont {J.~E.}\ \bibnamefont {Chesebro}}, \bibinfo {author} {\bibfnamefont {C.}~\bibnamefont {Heliot}}, \bibinfo {author} {\bibfnamefont {M.-C.}\ \bibnamefont {Ramel}}, \bibinfo {author} {\bibfnamefont {R.~K.}\ \bibnamefont {Stone}},\ and\ \bibinfo {author} {\bibfnamefont {C.~S.}\ \bibnamefont {Hill}},\ }\href@noop {} {\bibfield  {journal} {\bibinfo  {journal} {Developmental cell}\ }\textbf {\bibinfo {volume} {35}},\ \bibinfo {pages} {175} (\bibinfo {year} {2015})}\BibitemShut {NoStop}%
\bibitem [{\citenamefont {Jacobo}\ \emph {et~al.}(2019)\citenamefont {Jacobo}, \citenamefont {Dasgupta}, \citenamefont {Erzberger}, \citenamefont {Siletti},\ and\ \citenamefont {Hudspeth}}]{jacobo2019notch}%
  \BibitemOpen
  \bibfield  {author} {\bibinfo {author} {\bibfnamefont {A.}~\bibnamefont {Jacobo}}, \bibinfo {author} {\bibfnamefont {A.}~\bibnamefont {Dasgupta}}, \bibinfo {author} {\bibfnamefont {A.}~\bibnamefont {Erzberger}}, \bibinfo {author} {\bibfnamefont {K.}~\bibnamefont {Siletti}},\ and\ \bibinfo {author} {\bibfnamefont {A.}~\bibnamefont {Hudspeth}},\ }\href@noop {} {\bibfield  {journal} {\bibinfo  {journal} {Current Biology}\ }\textbf {\bibinfo {volume} {29}},\ \bibinfo {pages} {3579} (\bibinfo {year} {2019})}\BibitemShut {NoStop}%
\bibitem [{\citenamefont {Blackbeard}\ \emph {et~al.}(2012)\citenamefont {Blackbeard}, \citenamefont {Osborne}, \citenamefont {O'Brien},\ and\ \citenamefont {Amann}}]{blackbeard2012bursting}%
  \BibitemOpen
  \bibfield  {author} {\bibinfo {author} {\bibfnamefont {N.}~\bibnamefont {Blackbeard}}, \bibinfo {author} {\bibfnamefont {S.}~\bibnamefont {Osborne}}, \bibinfo {author} {\bibfnamefont {S.}~\bibnamefont {O'Brien}},\ and\ \bibinfo {author} {\bibfnamefont {A.}~\bibnamefont {Amann}},\ }\href@noop {} {\bibfield  {journal} {\bibinfo  {journal} {arXiv preprint arXiv:1210.4484}\ } (\bibinfo {year} {2012})}\BibitemShut {NoStop}%
\bibitem [{\citenamefont {Bormashenko}(2017)}]{bormashenko2017physics}%
  \BibitemOpen
  \bibfield  {author} {\bibinfo {author} {\bibfnamefont {E.~Y.}\ \bibnamefont {Bormashenko}},\ }\href@noop {} {\emph {\bibinfo {title} {Physics of wetting: phenomena and applications of fluids on surfaces}}}\ (\bibinfo  {publisher} {Walter de Gruyter GmbH \& Co KG},\ \bibinfo {year} {2017})\BibitemShut {NoStop}%
\bibitem [{\citenamefont {Boareto}\ \emph {et~al.}(2015)\citenamefont {Boareto}, \citenamefont {Jolly}, \citenamefont {Ben-Jacob},\ and\ \citenamefont {Onuchic}}]{boareto2015jagged}%
  \BibitemOpen
  \bibfield  {author} {\bibinfo {author} {\bibfnamefont {M.}~\bibnamefont {Boareto}}, \bibinfo {author} {\bibfnamefont {M.~K.}\ \bibnamefont {Jolly}}, \bibinfo {author} {\bibfnamefont {E.}~\bibnamefont {Ben-Jacob}},\ and\ \bibinfo {author} {\bibfnamefont {J.~N.}\ \bibnamefont {Onuchic}},\ }\href@noop {} {\bibfield  {journal} {\bibinfo  {journal} {Proceedings of the National Academy of Sciences}\ }\textbf {\bibinfo {volume} {112}},\ \bibinfo {pages} {E3836} (\bibinfo {year} {2015})}\BibitemShut {NoStop}%
\bibitem [{\citenamefont {Shi}\ and\ \citenamefont {Reimers}(2018)}]{shi2018understanding}%
  \BibitemOpen
  \bibfield  {author} {\bibinfo {author} {\bibfnamefont {X.}~\bibnamefont {Shi}}\ and\ \bibinfo {author} {\bibfnamefont {J.~R.}\ \bibnamefont {Reimers}},\ }\href@noop {} {\bibfield  {journal} {\bibinfo  {journal} {Scientific Reports}\ }\textbf {\bibinfo {volume} {8}},\ \bibinfo {pages} {2147} (\bibinfo {year} {2018})}\BibitemShut {NoStop}%
\bibitem [{\citenamefont {Dubrulle}\ \emph {et~al.}(2015)\citenamefont {Dubrulle}, \citenamefont {Jordan}, \citenamefont {Akhmetova}, \citenamefont {Farrell}, \citenamefont {Kim}, \citenamefont {Solnica-Krezel},\ and\ \citenamefont {Schier}}]{dubrulle2015response}%
  \BibitemOpen
  \bibfield  {author} {\bibinfo {author} {\bibfnamefont {J.}~\bibnamefont {Dubrulle}}, \bibinfo {author} {\bibfnamefont {B.~M.}\ \bibnamefont {Jordan}}, \bibinfo {author} {\bibfnamefont {L.}~\bibnamefont {Akhmetova}}, \bibinfo {author} {\bibfnamefont {J.~A.}\ \bibnamefont {Farrell}}, \bibinfo {author} {\bibfnamefont {S.-H.}\ \bibnamefont {Kim}}, \bibinfo {author} {\bibfnamefont {L.}~\bibnamefont {Solnica-Krezel}},\ and\ \bibinfo {author} {\bibfnamefont {A.~F.}\ \bibnamefont {Schier}},\ }\href@noop {} {\bibfield  {journal} {\bibinfo  {journal} {Elife}\ }\textbf {\bibinfo {volume} {4}},\ \bibinfo {pages} {e05042} (\bibinfo {year} {2015})}\BibitemShut {NoStop}%
\bibitem [{\citenamefont {Nandagopal}\ \emph {et~al.}(2018)\citenamefont {Nandagopal}, \citenamefont {Santat}, \citenamefont {LeBon}, \citenamefont {Sprinzak}, \citenamefont {Bronner},\ and\ \citenamefont {Elowitz}}]{nandagopal2018dynamic}%
  \BibitemOpen
  \bibfield  {author} {\bibinfo {author} {\bibfnamefont {N.}~\bibnamefont {Nandagopal}}, \bibinfo {author} {\bibfnamefont {L.~A.}\ \bibnamefont {Santat}}, \bibinfo {author} {\bibfnamefont {L.}~\bibnamefont {LeBon}}, \bibinfo {author} {\bibfnamefont {D.}~\bibnamefont {Sprinzak}}, \bibinfo {author} {\bibfnamefont {M.~E.}\ \bibnamefont {Bronner}},\ and\ \bibinfo {author} {\bibfnamefont {M.~B.}\ \bibnamefont {Elowitz}},\ }\href@noop {} {\bibfield  {journal} {\bibinfo  {journal} {Cell}\ }\textbf {\bibinfo {volume} {172}},\ \bibinfo {pages} {869} (\bibinfo {year} {2018})}\BibitemShut {NoStop}%
\bibitem [{\citenamefont {Viswanathan}\ \emph {et~al.}(2021)\citenamefont {Viswanathan}, \citenamefont {Hartmann}, \citenamefont {Pallares~Cartes},\ and\ \citenamefont {De~Renzis}}]{viswanathan2021desensitisation}%
  \BibitemOpen
  \bibfield  {author} {\bibinfo {author} {\bibfnamefont {R.}~\bibnamefont {Viswanathan}}, \bibinfo {author} {\bibfnamefont {J.}~\bibnamefont {Hartmann}}, \bibinfo {author} {\bibfnamefont {C.}~\bibnamefont {Pallares~Cartes}},\ and\ \bibinfo {author} {\bibfnamefont {S.}~\bibnamefont {De~Renzis}},\ }\href@noop {} {\bibfield  {journal} {\bibinfo  {journal} {The EMBO Journal}\ }\textbf {\bibinfo {volume} {40}},\ \bibinfo {pages} {e107245} (\bibinfo {year} {2021})}\BibitemShut {NoStop}%
\bibitem [{\citenamefont {Sonnen}\ and\ \citenamefont {Janda}(2021)}]{sonnen2021signalling}%
  \BibitemOpen
  \bibfield  {author} {\bibinfo {author} {\bibfnamefont {K.~F.}\ \bibnamefont {Sonnen}}\ and\ \bibinfo {author} {\bibfnamefont {C.~Y.}\ \bibnamefont {Janda}},\ }\href@noop {} {\bibfield  {journal} {\bibinfo  {journal} {Biochemical Journal}\ }\textbf {\bibinfo {volume} {478}},\ \bibinfo {pages} {4045} (\bibinfo {year} {2021})}\BibitemShut {NoStop}%
\bibitem [{\citenamefont {Casani-Galdon}\ and\ \citenamefont {Garcia-Ojalvo}(2022)}]{casani2022signaling}%
  \BibitemOpen
  \bibfield  {author} {\bibinfo {author} {\bibfnamefont {P.}~\bibnamefont {Casani-Galdon}}\ and\ \bibinfo {author} {\bibfnamefont {J.}~\bibnamefont {Garcia-Ojalvo}},\ }\href@noop {} {\bibfield  {journal} {\bibinfo  {journal} {Current Opinion in Cell Biology}\ }\textbf {\bibinfo {volume} {78}},\ \bibinfo {pages} {102130} (\bibinfo {year} {2022})}\BibitemShut {NoStop}%
\bibitem [{\citenamefont {Purvis}\ and\ \citenamefont {Lahav}(2013)}]{purvis2013encoding}%
  \BibitemOpen
  \bibfield  {author} {\bibinfo {author} {\bibfnamefont {J.~E.}\ \bibnamefont {Purvis}}\ and\ \bibinfo {author} {\bibfnamefont {G.}~\bibnamefont {Lahav}},\ }\href@noop {} {\bibfield  {journal} {\bibinfo  {journal} {Cell}\ }\textbf {\bibinfo {volume} {152}},\ \bibinfo {pages} {945} (\bibinfo {year} {2013})}\BibitemShut {NoStop}%
\bibitem [{\citenamefont {P{\'e}rez-Verdugo}\ \emph {et~al.}(2023)\citenamefont {P{\'e}rez-Verdugo}, \citenamefont {Banks},\ and\ \citenamefont {Banerjee}}]{perez2023excitable}%
  \BibitemOpen
  \bibfield  {author} {\bibinfo {author} {\bibfnamefont {F.}~\bibnamefont {P{\'e}rez-Verdugo}}, \bibinfo {author} {\bibfnamefont {S.}~\bibnamefont {Banks}},\ and\ \bibinfo {author} {\bibfnamefont {S.}~\bibnamefont {Banerjee}},\ }\href@noop {} {\bibfield  {journal} {\bibinfo  {journal} {arXiv preprint arXiv:2310.04950}\ } (\bibinfo {year} {2023})}\BibitemShut {NoStop}%
\bibitem [{\citenamefont {Bailles}\ \emph {et~al.}(2022)\citenamefont {Bailles}, \citenamefont {Gehrels},\ and\ \citenamefont {Lecuit}}]{bailles2022mechanochemical}%
  \BibitemOpen
  \bibfield  {author} {\bibinfo {author} {\bibfnamefont {A.}~\bibnamefont {Bailles}}, \bibinfo {author} {\bibfnamefont {E.~W.}\ \bibnamefont {Gehrels}},\ and\ \bibinfo {author} {\bibfnamefont {T.}~\bibnamefont {Lecuit}},\ }\href@noop {} {\bibfield  {journal} {\bibinfo  {journal} {Annual review of cell and developmental biology}\ }\textbf {\bibinfo {volume} {38}},\ \bibinfo {pages} {321} (\bibinfo {year} {2022})}\BibitemShut {NoStop}%
\bibitem [{\citenamefont {Di~Talia}\ and\ \citenamefont {Vergassola}(2022)}]{di2022waves}%
  \BibitemOpen
  \bibfield  {author} {\bibinfo {author} {\bibfnamefont {S.}~\bibnamefont {Di~Talia}}\ and\ \bibinfo {author} {\bibfnamefont {M.}~\bibnamefont {Vergassola}},\ }\href@noop {} {\bibfield  {journal} {\bibinfo  {journal} {Annual review of biophysics}\ }\textbf {\bibinfo {volume} {51}},\ \bibinfo {pages} {327} (\bibinfo {year} {2022})}\BibitemShut {NoStop}%
\bibitem [{\citenamefont {Corominas-Murtra}\ and\ \citenamefont {Petridou}(2021)}]{corominas2021viscoelastic}%
  \BibitemOpen
  \bibfield  {author} {\bibinfo {author} {\bibfnamefont {B.}~\bibnamefont {Corominas-Murtra}}\ and\ \bibinfo {author} {\bibfnamefont {N.~I.}\ \bibnamefont {Petridou}},\ }\href@noop {} {\bibfield  {journal} {\bibinfo  {journal} {Frontiers in Physics}\ }\textbf {\bibinfo {volume} {9}},\ \bibinfo {pages} {666916} (\bibinfo {year} {2021})}\BibitemShut {NoStop}%
\bibitem [{\citenamefont {Guirao}\ and\ \citenamefont {Bella{\"\i}che}(2017)}]{guirao2017biomechanics}%
  \BibitemOpen
  \bibfield  {author} {\bibinfo {author} {\bibfnamefont {B.}~\bibnamefont {Guirao}}\ and\ \bibinfo {author} {\bibfnamefont {Y.}~\bibnamefont {Bella{\"\i}che}},\ }\href@noop {} {\bibfield  {journal} {\bibinfo  {journal} {Current opinion in cell biology}\ }\textbf {\bibinfo {volume} {48}},\ \bibinfo {pages} {113} (\bibinfo {year} {2017})}\BibitemShut {NoStop}%
\bibitem [{\citenamefont {Hill}(2018)}]{hill2018spatial}%
  \BibitemOpen
  \bibfield  {author} {\bibinfo {author} {\bibfnamefont {C.~S.}\ \bibnamefont {Hill}},\ }\href@noop {} {\bibfield  {journal} {\bibinfo  {journal} {Current opinion in cell biology}\ }\textbf {\bibinfo {volume} {51}},\ \bibinfo {pages} {50} (\bibinfo {year} {2018})}\BibitemShut {NoStop}%
\bibitem [{\citenamefont {Efron}(1982)}]{efron1982jackknife}%
  \BibitemOpen
  \bibfield  {author} {\bibinfo {author} {\bibfnamefont {B.}~\bibnamefont {Efron}},\ }\href@noop {} {\emph {\bibinfo {title} {The jackknife, the bootstrap and other resampling plans}}}\ (\bibinfo  {publisher} {SIAM},\ \bibinfo {year} {1982})\BibitemShut {NoStop}%
\bibitem [{\citenamefont {Heisenberg}\ \emph {et~al.}(2000)\citenamefont {Heisenberg}, \citenamefont {Tada}, \citenamefont {Rauch}, \citenamefont {Sa{\'u}de}, \citenamefont {Concha}, \citenamefont {Geisler}, \citenamefont {Stemple}, \citenamefont {Smith},\ and\ \citenamefont {Wilson}}]{heisenberg2000silberblick}%
  \BibitemOpen
  \bibfield  {author} {\bibinfo {author} {\bibfnamefont {C.-P.}\ \bibnamefont {Heisenberg}}, \bibinfo {author} {\bibfnamefont {M.}~\bibnamefont {Tada}}, \bibinfo {author} {\bibfnamefont {G.-J.}\ \bibnamefont {Rauch}}, \bibinfo {author} {\bibfnamefont {L.}~\bibnamefont {Sa{\'u}de}}, \bibinfo {author} {\bibfnamefont {M.~L.}\ \bibnamefont {Concha}}, \bibinfo {author} {\bibfnamefont {R.}~\bibnamefont {Geisler}}, \bibinfo {author} {\bibfnamefont {D.~L.}\ \bibnamefont {Stemple}}, \bibinfo {author} {\bibfnamefont {J.~C.}\ \bibnamefont {Smith}},\ and\ \bibinfo {author} {\bibfnamefont {S.~W.}\ \bibnamefont {Wilson}},\ }\href@noop {} {\bibfield  {journal} {\bibinfo  {journal} {Nature}\ }\textbf {\bibinfo {volume} {405}},\ \bibinfo {pages} {76} (\bibinfo {year} {2000})}\BibitemShut {NoStop}%
\bibitem [{\citenamefont {Keller}\ \emph {et~al.}(2008)\citenamefont {Keller}, \citenamefont {Schmidt}, \citenamefont {Wittbrodt},\ and\ \citenamefont {Stelzer}}]{keller2008reconstruction}%
  \BibitemOpen
  \bibfield  {author} {\bibinfo {author} {\bibfnamefont {P.~J.}\ \bibnamefont {Keller}}, \bibinfo {author} {\bibfnamefont {A.~D.}\ \bibnamefont {Schmidt}}, \bibinfo {author} {\bibfnamefont {J.}~\bibnamefont {Wittbrodt}},\ and\ \bibinfo {author} {\bibfnamefont {E.~H.}\ \bibnamefont {Stelzer}},\ }\href@noop {} {\bibfield  {journal} {\bibinfo  {journal} {science}\ }\textbf {\bibinfo {volume} {322}},\ \bibinfo {pages} {1065} (\bibinfo {year} {2008})}\BibitemShut {NoStop}%
\bibitem [{\citenamefont {Schindelin}\ \emph {et~al.}(2012)\citenamefont {Schindelin}, \citenamefont {Arganda-Carreras}, \citenamefont {Frise}, \citenamefont {Kaynig}, \citenamefont {Longair}, \citenamefont {Pietzsch}, \citenamefont {Preibisch}, \citenamefont {Rueden}, \citenamefont {Saalfeld}, \citenamefont {Schmid} \emph {et~al.}}]{schindelin2012fiji}%
  \BibitemOpen
  \bibfield  {author} {\bibinfo {author} {\bibfnamefont {J.}~\bibnamefont {Schindelin}}, \bibinfo {author} {\bibfnamefont {I.}~\bibnamefont {Arganda-Carreras}}, \bibinfo {author} {\bibfnamefont {E.}~\bibnamefont {Frise}}, \bibinfo {author} {\bibfnamefont {V.}~\bibnamefont {Kaynig}}, \bibinfo {author} {\bibfnamefont {M.}~\bibnamefont {Longair}}, \bibinfo {author} {\bibfnamefont {T.}~\bibnamefont {Pietzsch}}, \bibinfo {author} {\bibfnamefont {S.}~\bibnamefont {Preibisch}}, \bibinfo {author} {\bibfnamefont {C.}~\bibnamefont {Rueden}}, \bibinfo {author} {\bibfnamefont {S.}~\bibnamefont {Saalfeld}}, \bibinfo {author} {\bibfnamefont {B.}~\bibnamefont {Schmid}}, \emph {et~al.},\ }\href@noop {} {\bibfield  {journal} {\bibinfo  {journal} {Nature methods}\ }\textbf {\bibinfo {volume} {9}},\ \bibinfo {pages} {676} (\bibinfo {year} {2012})}\BibitemShut {NoStop}%
\bibitem [{\citenamefont {Dhooge}\ \emph {et~al.}(2008)\citenamefont {Dhooge}, \citenamefont {Govaerts}, \citenamefont {Kuznetsov}, \citenamefont {Meijer},\ and\ \citenamefont {Sautois}}]{dhooge2008new}%
  \BibitemOpen
  \bibfield  {author} {\bibinfo {author} {\bibfnamefont {A.}~\bibnamefont {Dhooge}}, \bibinfo {author} {\bibfnamefont {W.}~\bibnamefont {Govaerts}}, \bibinfo {author} {\bibfnamefont {Y.~A.}\ \bibnamefont {Kuznetsov}}, \bibinfo {author} {\bibfnamefont {H.~G.~E.}\ \bibnamefont {Meijer}},\ and\ \bibinfo {author} {\bibfnamefont {B.}~\bibnamefont {Sautois}},\ }\href@noop {} {\bibfield  {journal} {\bibinfo  {journal} {Mathematical and Computer Modelling of Dynamical Systems}\ }\textbf {\bibinfo {volume} {14}},\ \bibinfo {pages} {147} (\bibinfo {year} {2008})}\BibitemShut {NoStop}%
\bibitem [{\citenamefont {Inc.}()}]{Mathematica}%
  \BibitemOpen
  \bibfield  {author} {\bibinfo {author} {\bibfnamefont {W.~R.}\ \bibnamefont {Inc.}},\ }\href {https://www.wolfram.com/mathematica} {\bibinfo {title} {Mathematica, {V}ersion 13.0}},\ \bibinfo {note} {champaign, IL, 2021}\BibitemShut {NoStop}%
\end{thebibliography}%

% Supplementary Material
%-------------------------------
\section*{Supplementary Material}

\subsection{Contact areas of wet foams}
\label{sec:wet_foams}

\begin{figure}
    \centering
    \includegraphics[]{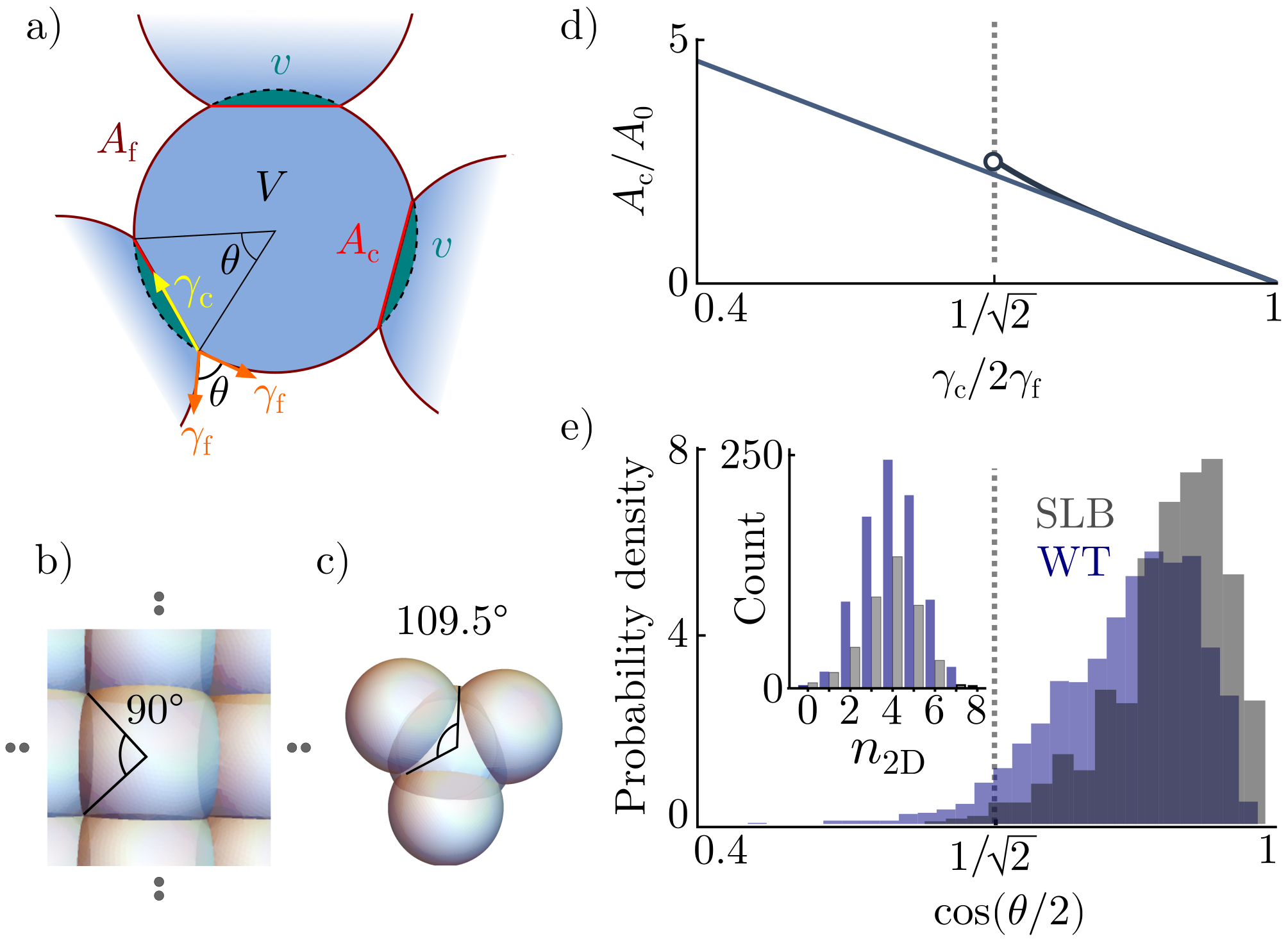}
    \caption{Contact areas in wet foams. (a) We consider droplets with volume $V$ in fixed-topology configurations with $n$ neighbors in the small contact angle limit (no higher-order junctions), with $A_{\rm{c}}$ the total contact area per droplet, and $A_{\rm{f}}$ the free surface area. The balance of surface tensions at the contact and free surfaces $\gamma_{\rm{c}},\gamma_{\rm{f}}$ defines the equilibrium contact angle $\theta$ [Eq.~\ref{eq:tension_angle}. (b) Square ($n=4$) or cubic ($n=6$) lattices and (c) five droplets in a tetrahedral configuration are shown at the onset of higher-order junction formation. (d)  Linearization of Eq.\ref{eq:contact_area} (solid curve, $n=6$) around $\gamma_{\rm{c}}/2\gamma_{\rm{f}}=1$ (dashed line) provides a good approximation for small contact angles, where the tension ratios are near 1 [Eq.\eqref{eq:contact_area_linear}]. (e) Contact angle measurements in zebrafish embryos show that the data is well described by the small angle limit for cubic droplet configurations ($95.22\%$ of WT and $98.75\%$ of SLB data points are above $1/\sqrt{2}$, the threshold for higher order junction formation). Inset: Histogram of the number of contacts per cell measured from 2D microscopy images $n_{\rm{2D}}$. WT: $N=871$, SLB: $N=429$}.
    \label{fig:wet_foam_contacts}
\end{figure}

We consider equilibrium droplet configurations with fixed topology, where each droplet with a fixed volume $V$ forms contacts with $n$ neighboring droplets without triple or higher-order junctions [Fig.\ref{fig:wet_foam_contacts}(a)]. The ratio of the uniform surface tensions $\gamma_{\rm{c}}$ at droplet-droplet interfaces and $\gamma_{\rm{f}}$ at the free surface determine the contact angle $\theta$ [Fig.~\ref{fig:wet_foam_contacts}(a)] through the force balance equation
\begin{align}
    \frac{\gamma_{\rm{c}}}{2\gamma_{\rm{f}}} = \cos\frac{\theta}{2}
    \label{eq:tension_angle}
\end{align}
In the minimal surface configuration, the droplets take the shape of truncated spheres from which $n$ identical spherical caps were removed [Fig.\ref{fig:wet_foam_contacts}(a)], each with a volume
\begin{align}
    v = \frac{\pi}{3}r^3 \left(
    2+\cos\frac{\theta}{2} 
    \right)\left( 
    1-\cos\frac{\theta}{2}
    \right)^2,
    \label{eq:v}
\end{align}
in which $r$ is the radius of the contact.
The total contact area $A_{\rm{c}}=n \pi r^2$ can be related to the droplet volume $V$ through
\begin{align}
    A_{\rm{c}} = n \left[
    1-\cos^2\frac{\theta}{2}
    \right]
    \left[\frac{
    3 \pi^{1/2} (V + (n-1) v))
    }{
    \left( 2-\cos\dfrac{\theta}{2} \right) 
    \left( 1 + \cos\dfrac{\theta}{2} \right)^2
    }\right]^{2/3},
    \label{eq:AcCap}
\end{align}
using that $V + (n-1) v$ corresponds to the volume of a spherical cap. From Eqs.~\eqref{eq:tension_angle}--\eqref{eq:AcCap} follows Eq.~\eqref{eq:contact_area}, which is valid until the formation of triple or higher order junctions, e.g. at $\theta=90^{\circ}$ for square $(n=4)$ and cubic $(n=6)$ lattice topologies [Fig.~\ref{fig:Fig1}(a) and Fig.~\ref{fig:wet_foam_contacts}(b)].  In the case of droplets with four equally spaced contacts--forming a tetrahedral configuration--the corresponding angle is $\theta=109.5^{\circ}$ [Fig.~\ref{fig:wet_foam_contacts}(c)], however, tetrahedral arrangements are not space-filling in three dimensions.

\subsection{Modeling contact angles in zebrafish embryos}
\label{sec:contact_angle_modeling}
\subsubsection{Overview} 

Mesendoderm formation in zebrafish embryos is an early developmental event in which the cells that later form the organism's internal organs differentiate and alter their material properties. This process is guided by a spatial gradient of Nodal signaling activity, which decreases along the animal-to-vegetal embryo axis (AV-axis) from the margin towards the animal pole [Fig.~\ref{fig:embryo}(b)] \cite{hill2018spatial}, and involves changes in cell-cell adhesion \cite{petridou2021rigidity}.
We obtained fluorescence imaging data of wild type (WT) and \emph{silberblick} mutant (SLB) embryos 5\,h after fertilization, and measured the contact angles $\theta_j(y_j)$ between cells of the blastoderm at different positions $y_j$ along the AV-axis from the margin at $y = 0$. We obtained 
$N=2132$ contact angle datapoints from five WT embryos, and $N=806$ datapoints from three SLB embryos.
We then used Eq.~\eqref{eq:tension_angle} to relate the measured contact angles to the steady state tension ratios obtained from simulating our equations (see below for details). In particular, we use simulation-based inference (SBI) to infer parameter describing the distributions $p_{\rm{WT}} (\cos(\theta/2))$ and $p_{\rm{SLB}} (\cos(\theta/2))$ shown in Fig.\ref{fig:embryo}(b).

\subsubsection{Modeling tissues as wet-limit foams with fixed topology in the small contact angle limit}
We measured the number of in-plane contacts per cell across the blastoderm from 2D microscopy images (as described in \cite{petridou2021rigidity}), and obtained $4.05\pm0.05$ in the WT and $3.77\pm0.07$ in the SLB mutant [Fig.~\ref{fig:wet_foam_contacts}(e)](mean$\pm$standard error). Extrapolating from these in-plane measurements suggests that blastoderm cells have an average of six neighbors in three dimensions, and that the system is close to the rigidity percolation threshold \cite{petridou2021rigidity}. Therefore, we model the non-confluent 3D blastoderm tissue as a fixed-topology configuration of droplets with $n=6$ contacts, which corresponds to the contact number of disordered wet-limit foams close to the jamming/unjamming transition \cite{drenckhan2015structure}. 
For the cubic lattice, higher order junctions form at contact angles $\theta \geq 90^{\circ}$ [Fig.~\ref{fig:wet_foam_contacts}(b)], corresponding to $\cos\left(\theta/2\right)<1/\sqrt{2}$. Our contact angle measurements show that more than $95\%$ of all data points fall above this point [Fig.~\ref{fig:wet_foam_contacts}(d, e)]. In this small-angle regime, contact areas are well approximated by linearizing Eq.~\eqref{eq:contact_area} around the tension ratio at detachment  $\gamma_{\rm{c}}/2\gamma_{\rm{f}}=1$ [Fig.~\ref{fig:wet_foam_contacts}(d)]
\begin{align}
    \frac{A_{\rm{c}}}{A_0} = n 2^{1/3}\left(1-\frac{\gamma_{\rm{c}}}{2\gamma_{\rm{f}}} \right)+ \mathcal{O}\left(\left(
    1 -  \frac{\gamma_{\rm{c}}}{2\gamma_{\rm{f}}} 
    \right)^2 \right),
    \label{eq:contact_area_linear}
\end{align}
which we use for the parameter estimations.

\subsubsection{Cell state dynamics are governed by an external signaling gradient}
The blastoderm cells respond to extracellular Nodal signals \cite{barone2017effective,muller2012differential}, which we model as an exponentially decaying stochastic concentration field [Fig.~\ref{fig:embryo}(b)] \cite{muller2012differential} $\phi(y) = \langle \phi(y) \rangle + \eta(y)$ including a Poissonian noise term to account for molecule fluctuations \cite[Appendix III]{chandrasekhar1943stochastic}. For each position $y_j$ at which a contact-angle measurement is available, we solve Eqs.~\eqref{eq:u}--\eqref{eq:signal} in the local approximation, i.e. 
\begin{align}
    \tau_u \frac{du}{dt} = \frac{
    \left( \chi \phi(y_j) \dfrac{A_{\rm{c}}}{A_0} \right)^h
    }{
    1+\left( \chi \phi(y_j) \dfrac{A_{\rm{c}}}{A_0} \right)^h
    } - u,
    \label{eq:ode_foam}
\end{align}
whereby we neglect differences in the external signal received by neighboring cells, and nearest-neighbor variations in the contact areas. 
The steady states of Eq.~\eqref{eq:ode_foam} depend on the parameters $\gamma_0/2\gamma_{\rm{f}}$, $\gamma_{\rm{A}}/\gamma_0$, the product $\chi\phi_0$, and the Hill coefficient $h$. 

\subsubsection{Simulation-based inference analysis}
\label{sec:sbi}
%In this section, we describe the methodology and implementation of
We used simulation-based inference (SBI) to infer the unknown parameters from the statistics of the measured contact angles across positions and samples [Fig.~\ref{fig:embryo}(b)]. SBI is particularly suitable for scenarios where the likelihood function is intractable or difficult to compute, but where simulating data from the model is straightforward. Given a set of observations $\mathbf{x}_{\rm{obs}}$ (here the summary statistics of measured contact angles, see below), SBI relies on Bayes' theorem for the probable set of parameters $\vartheta$ describing the data. In particular, we are interested in the posterior distribution
\begin{align}
    p(\vartheta | \mathbf{x_{\rm{obs}})} = \frac{
    p(\vartheta) p(\mathbf{x_{\rm{obs}}} | \vartheta)
    }{
    p(\mathbf{x_{\rm{obs}}})
    }
\end{align}
with $p(\vartheta)$ the prior over the parameters, and $p(\mathbf{x_{\rm{obs}}} | \vartheta)$ probed by stochastic simulations.

\paragraph{Simulation step.}
\label{sec:sbi_model}

Given a set of model parameter values (Hill coefficient $h$, the ratios of tension coefficients $\gamma_0/ 2 \gamma_{\rm f}$ and $\gamma_{\rm A} / \gamma_0$, and the product $\chi \phi_0$), our simulator evaluates the steady state of Eq.~\eqref{eq:ode_foam} using \eqref{eq:tension_regulation} and \eqref{eq:contact_area_linear}, starting from random initial conditions $u(t=0)\in[0,1]$ using \emph{solve\_ivp} with the \emph{RK45}-method (explicit fourth order Runge-Kutta) from the \emph{scipy} python package. For each position $y$ at which a contact angle was measured, one simulation was performed. To account for the Poissonian statistics of fluctuating concentrations in the external signal gradient, $\chi \phi$ was drawn from a $\Gamma$-distribution with mean $\mu_{\rm{\chi\phi}} = \langle \chi \phi(y) \rangle$ and variance $\sigma_{\rm{\chi\phi}} = \langle \chi \phi(y) \rangle$.
We used [Eq.~\eqref{eq:tension_angle}] to calculate contact angles from steady state tension ratios and added a relative error of $15\%$ to account for experimental measurement errors by drawing a new contact angle from a normal distribution centered at $\mu_{\theta}$ equal to the simulated steady state angle and standard deviation $\sigma_{\rm{\theta}} = 0.15 \mu_{\rm{\theta}}$.

Our preliminary analysis revealed that in contrast to the tension and susceptibility parameters, 
the Hill coefficient was not well constrained by the data, but the shape of the inferred posterior suggests values of $h>2$ to be most suitable, with a broad peak around $h=7$ [Fig.~\ref{fig:sbi_hill_analysis}(a)]. Following common convention for biological signaling models, we therefore fixed the Hill coefficient to $h=4$ \cite{boareto2015jagged, shi2018understanding, dubrulle2015response}.

There remain three parameters $\vartheta = (\gamma_0/2\gamma_{\rm{f}}, \gamma_{\rm{A}}/\gamma_0, \chi\phi_0)$, which can be identified from the measurements and which are assumed distributed with uniform priors
\begin{align}
    p(\gamma_0/2\gamma_{\rm{f}}) &= \mathcal{U}([0.7, 1]), \nonumber \\
    p(\gamma_{\rm{A}}/\gamma_0) &= \mathcal{U}([0,1]), \label{eq:prior} \\
    p(\chi\phi_0) &= \mathcal{U}([0,20]), \nonumber
\end{align}
where $\mathcal{U}([a,b])$ denotes a uniform distribution on the interval $[a,b]$. The range of baseline tension ratios $\gamma_0/2\gamma_{\rm{f}}$ was chosen such that---considering Eqs.~\eqref{eq:tension_angle},\eqref{eq:tension_regulation} and $\gamma_{\rm{A}}=0$---the corresponding values of contact angles cover the range of measurements. We also tested that a broader prior $p(\gamma_0/2\gamma_{\rm{f}}) = \mathcal{U}([0, 1])$ yields the same results, but considerably slows down the inference procedure. The adaptive tension ratio $\gamma_A/\gamma_0$ was sampled from the full domain in which the theory is valid.
For $\chi\phi_0$ we chose an upper limit of 20 (an order of magnitude above $(\chi\phi_0)_{\rm{Cusp}}$, [Fig.~\ref{fig:embryo}(a),(c)]), however, the same results were obtained with a larger prior range of $p(\chi\phi_0) = \mathcal{U}([0,100])$.

For the genetically perturbed embryos (SLB mutant), $\gamma_0/2\gamma_{\rm{f}}$ and $\chi\phi_0$ were fixed to the values inferred from the wildtype data, leaving $\gamma_{\rm{A}}/\gamma_{0}$ as the only free parameter.

\begin{figure}
    \centering
    \includegraphics[width=\linewidth]{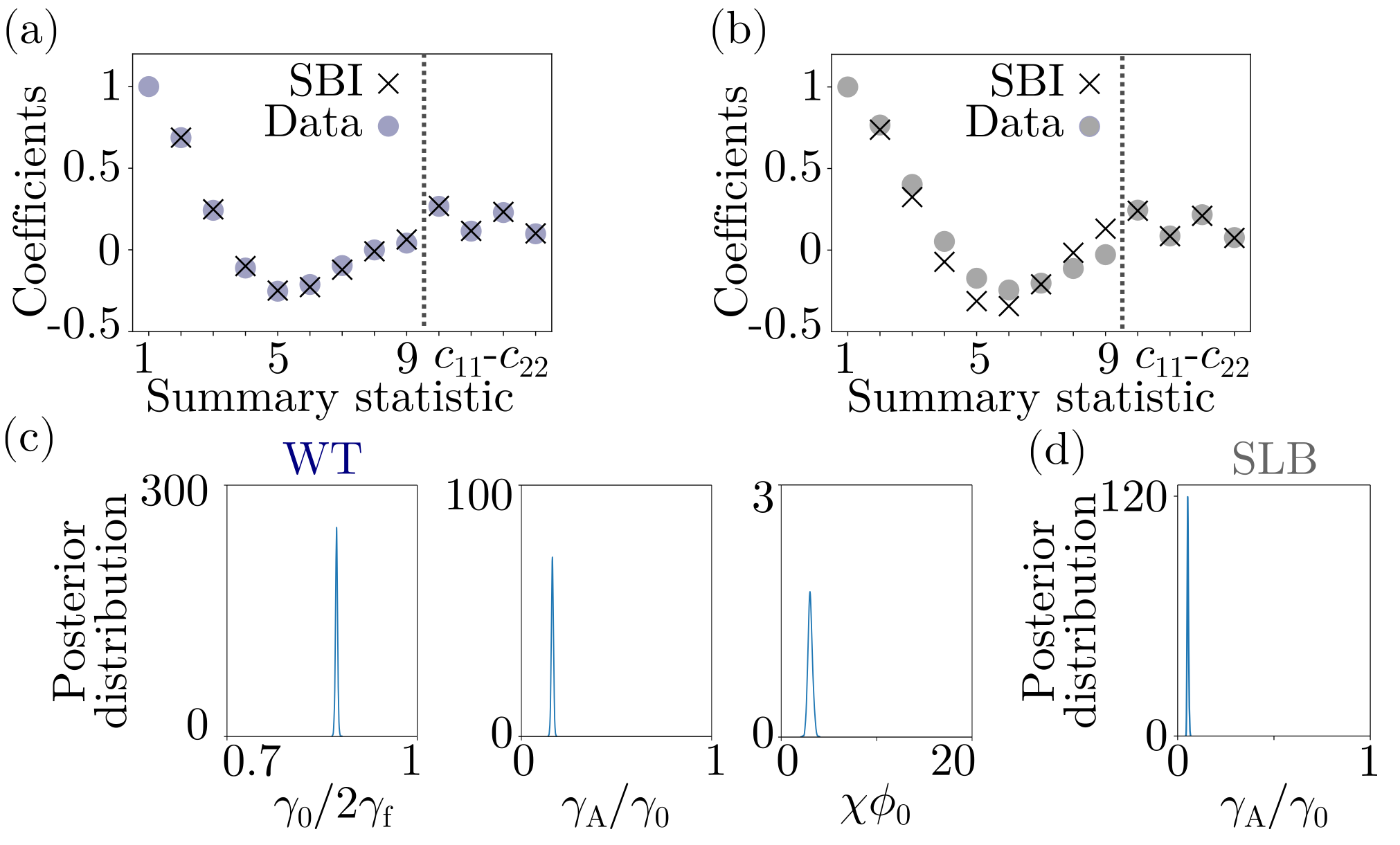}
    \caption{SBI analysis of contact angle distributions (a,b) Data: Legendre and cross-moment coefficients [Eq.~\eqref{eq:corr_coeff}] of the measured distribution of $\cos\bm{(}\theta(y)/2\bm{)}$ [Fig.\ref{fig:embryo}(b)], SBI: Inferred parameters were used to simulate distributions of $\cos\bm{(}\theta(y)/2\bm{)}$, from which Legendre and cross-moment coefficients were computed. Error bars representing the standard deviation of the posterior are too small to be displayed due to the narrow posterior distributions (compare to c,d). (c,d) Distribution of parameter predictions from sampling the trained posterior 2000 times for the wild type (c) and SLB mutant data (d).}
    \label{fig:fit_statistics}
\end{figure}

\begin{figure}
    \centering
    \includegraphics[width=\linewidth]{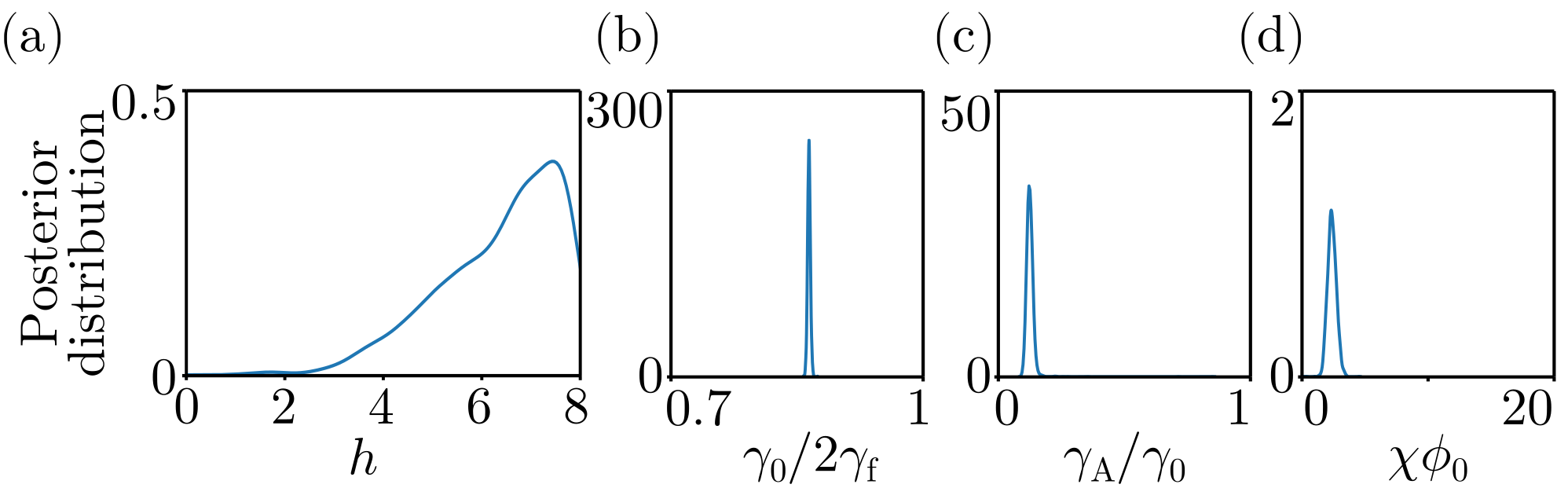}
    \caption{Posterior distributions of four model parameters inferred using simulation-based inference (SBI) on WT data: (a) Hill coefficient $h$, (b) ratio $\gamma_0/2\gamma_{\rm{f}}$ (c) ratio $\gamma_{\rm{A}}/\gamma_0$, and (d) $\chi \phi_0$. While parameters (b–d) are well constrained by the data ($\gamma_0/2\gamma_{\rm{f}}=0.864\pm0.002$, $\gamma_{\rm{A}}/\gamma_0=0.13\pm0.03$, $\chi \phi_0 = 2.3\pm0.4$, errors are the standard deviations of the posterior), the Hill coefficient $h$ remains poorly constrained, though the analysis suggests that $h>2$ best describes the system. Distributions were obtained by sampling the posterior 2000 times. We used priors as given in Eq.~\eqref{eq:prior} for parameters (b-d) and a prior of $p(h)=\mathcal{U}([0,8])$ for the Hill coefficient, otherwise SBI analysis was performed as described in Sec.~\ref{sec:sbi}}
    \label{fig:sbi_hill_analysis}
\end{figure}

\paragraph{Training}
\label{sec:sbi_implementation}
As the summary statics $\bm{x}_{\rm obs}$---the features extracted from the measurements or results of simulations---we used
the moments
$$
    \ell_{m} = \langle\mathcal{L}_m\bm{(}\cos(\theta / 2)\bm{)}\rangle
$$
of the \textit{shifted Legendre polynomials} $\mathcal{L}_m$ of orders $m = 1, 2, ... 8$ (order $m = 0$ yields 1 due to the normalization of the probability density), which characterize the marginal distribution of the contact angles $p_{\rm{WT}}\bm{(}\cos(\theta/2)\bm{)}$ and $p_{\rm{SLB}}\bm{(}\cos(\theta/2)\bm{)}$ [Fig.~\ref{fig:fit_statistics}(a,b)].

To include information about the spatial structure in the data, we additionally computed four cross-moments
\begin{align}
    c_{\alpha \beta} = \frac{1}{N}\sum_{j=1}^N y_j^\alpha \cos^\beta \left[\frac{\theta(y_j)}{2} \right],
    \label{eq:corr_coeff}
\end{align}
with $\alpha,\beta\in\{1,2\}$ and $N$ being the number of data points. 

In total we thus obtained twelve degrees of freedom $\bm{x}_{\rm obs}$(8 Legendre moments and 4 cross-moments), which we used to train the posterior estimator $p(\vartheta | \bm{x}_{\rm obs})$. To this end we leveraged the python implementation of the SBI method~\cite{Tejero-Cantero2020}. In particular, we used the sequential neural posterior estimator (SNPE) with the neural-spline flow representation of distribution functions.

The training set included $5\times10^5$ simulations of wildtype embryos with three variable parameters sampled from Eq.~\eqref{eq:prior}, and $10^5$ simulations of the \emph{silberblick} embryos. The expected values of the inferred parameters were calculated over 2000 samples from the obtained posterior distribution $p(\vartheta|\bm{x}_{\rm obs})$ [Fig.~\ref{fig:fit_statistics}(c-d)].

To assess the error of the parameter inference arising from sample-to-sample variability between different embryos (reported in the main text and Fig.~\ref{fig:embryo}(c)), we used cross-validation. Specifically, we computed the standard error of inferred parameter $\vartheta$ using \textit{jackknife resampling}~\cite{efron1982jackknife}
\begin{align}
    \text{std}_{\vartheta} = \sqrt{
    \frac{1}{M (M-1)} \sum_{k=1}^M \left( 
    \vartheta_{\text{k}} - \langle \vartheta \rangle 
    \right)^2 
    },
    \label{eq:jackknife}
\end{align}
where $M$ is the number of embryos, $\vartheta_{\rm{k}}$ is the inferred parameter value obtained using all but the data from the $k$-th embryo for training the neural network to estimate the posterior, and $\langle \vartheta \rangle$ is the mean of the $M$ different inferred parameter values.

\subsection{Experimental methods}

\subsubsection{Fish Raising and embryo collection}

Control (AB2B2) and wnt11/slb f2 mutant \cite{heisenberg2000silberblick} Zebrafish were maintained as described in \cite{petridou2021rigidity}. Embryos were collected and raised in E3 and Danieau's media at $28.5^{\circ}C$.

\subsubsection{Embryo micro-injections}

Embryos were injected at 1-cell stage (0 hours post fertilization (hpf)) with \SI{70}{\pico\gram} membrane-RFP (Iioka 20024) mRNA and \SI{70}{\pico\gram} histone-GFP \cite{keller2008reconstruction} for membrane and nuclei fluorescent labelling. At high stage (3 hpf) the blastoderm was injected with \SI{1}{\nano\litre} of \SI{0.6}{\milli\gram\per\milli\litre} dextran Alexa Fluor 647 (10'000MW; D22914, Invitrogen) to label the interstitial fluid.

\subsubsection{Image acquisition}

Live imaging was performed on an upright Zeiss LSM980 equipped with a 20x objective (W Plan-Apochromat water immersion objective). Dechorionated and injected embryos were mounted on customised Agarose moulds in petri dishes and immobilized with 0.5\% low melting point agarose (16520050 Invitrogen). Z-stack imaging was performed at regular time intervals of about \SI{10}{\minute}, from 3 to 7 hpf, with a Z interval of \SIrange{2}{3}{\micro\meter}.

\subsubsection{Analysis}

Cell-cell external contact angles ($\theta$) were measured with the FIJI angle tool \cite{schindelin2012fiji}, and the location in the blastoderm of each angle was determined as the distance from the YSL membrane ($y=0$).

\subsection{Numerical methods}
\label{sec:numerical_methods}

\begin{table*}[ht!]
    \centering
    \begin{tabular}{|p{70 mm}|p{20 mm}|p{60mm}|}
        \hline
         Physical quantity & Symbol & Values\\
         \hline
         Baseline interfacial tension relative to outer surface tension & $\gamma_0/2\gamma_{\rm{f}}$ & 
         Fig.~\ref{fig:embryo}(a): 0.87 \hfill \newline
          Fig.~\ref{fig:symmetric_doublet_results}(a-c): 0.98 \hfill \newline
         Fig.~\ref{fig:symmetric_doublet_results}(d): 0.95 \\
         \hline
         Critical susceptibility without adaptive tension & $\chi_0$ & Fig.~\ref{fig:symmetric_doublet_results},\ref{fig:SPDetails}:40.604 \\
         \hline
         Signal amplitude &  $\phi_0$ & Fig.~\ref{fig:symmetric_doublet_results},\ref{fig:SPDetails}: 1 \\
         \hline
         Adaptive adhesion coefficient relative to outer surface tension & $\gamma_{\rm{A}}/2\gamma_{\rm{f}}$ &
         Fig.~\ref{fig:embryo}(a) inlet: 0.261 \hfill \newline
         Fig.~\ref{fig:embryo}(d): 0.1392 \hfill \newline
         Fig.~\ref{fig:symmetric_doublet_results}(b): \{square, triangle: 0.15, quartrefoil:0.21, star:0.23, cross:0.2352, pentagon:0.5\} \hfill \newline
         Fig.~\ref{fig:symmetric_doublet_results}(d): 0.784 \hfill \newline 
        \\
         % Fig.~\ref{fig:asymmetric_doublets_results}(c): 0.65 \\
         \hline
         Relative signal susceptibility & $\chi/\chi_0$ &
         Fig.~\ref{fig:symmetric_doublet_results}(c): \{square:0.1, quartrefoil:0.61, star:0.604, cross:0.6021, pentagon:0.6, triangle:0.95\} \hfill \newline
         Fig.~\ref{fig:symmetric_doublet_results}(d): \{0.4704, 0.7388\} \\ 
         \hline
         Hill coefficient & h & 4 \\
         \hline
    \end{tabular}
    \caption{Parameter values.}
    \label{tab:parameter_values_figures}
\end{table*}

\begin{figure*}[ht!]
    \centering
    \includegraphics[width=17.8cm]{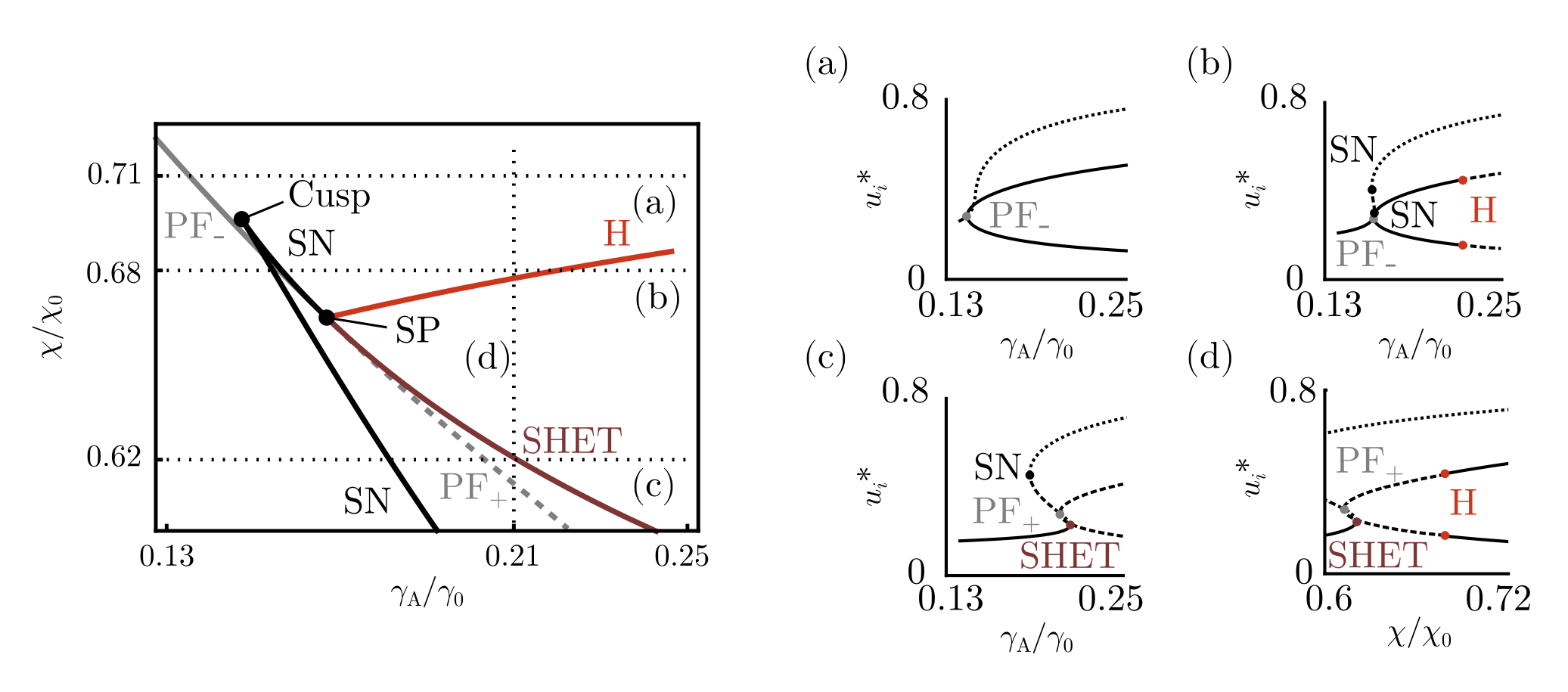}
    \caption{Bifurcation analysis close to the saddle-node pitchfork, 
    Left: Enlarged view of the state diagram of the doublet in terms of normalized feedback control parameters shown in Fig.~\ref{fig:symmetric_doublet_results}(a) close to the saddle-node pitchfork (SP) codimension-2 bifurcation point.  (a) - (d) show stable (solid line) and unstable (dashed line) fixpoints and saddles (dotted line) computed for variation of one feedback parameter as indicated by dotted lines in the state diagram on the left.  As the pitchfork interacts with one of the saddle-nodes (compare (b) and (c)), it changes from supercritical (PF$_-$) to subcritical (PF$_+$) and the saddle (SN) becomes a Saddle-Heteroclinic (SHET). In the parameter regime between the H and SHET bifurcation lines, the system has no stable fixpoints, but stable limit cycles. H: Hopf bifurcation. Diagrams were computed in MatCont (Sec.~\ref{sec:numerical_methods}). This result is also presented in the Appendix of the companion article~\cite[Fig.~11]{PREJoint}}
    \label{fig:SPDetails}
\end{figure*}

\paragraph{Bifurcation analysis}
The state and bifurcation diagrams presented in Fig.~\ref{fig:embryo}(a), Fig.~\ref{fig:symmetric_doublet_results}(a,d), and Fig.~\ref{fig:SPDetails} were computed via continuation with the MATLAB-based software package MatCont \cite{dhooge2008new} (MatCont7p4 and MATLAB R2021a, scripts with details and numerical settings available at https://git.embl.de/dullwebe/dullweber2024). In general, fixpoints to initialize the continuation were computed by integration over time using the Integrator Method ode45. 

Results of the continuation were confirmed using simulations and analysis in Mathematica 13.0~\cite{Mathematica} (notebook with a step-by-step explanation of the analysis available at https://git.embl.de/dullwebe/dullweber2024). Specifically, we tested the number and types of stable attractors in different parameter regimes with simulations using NDSolve and ParametricNDSolve with the equation simplification method \emph{Residuals}. Fixpoints shown in the phase plots Fig.~\ref{fig:symmetric_doublet_results}(b) were computed numerically in Mathematica from the intersections of nullclines. The oscillation amplitude [Fig.~\ref{fig:symmetric_doublet_results}(d),(e)] and period  [Fig.~\ref{fig:symmetric_doublet_results}(a)] were computed from the extrema of simulated trajectories, and checked against the dominant Fourier components.\\

\paragraph{Numerical surface energy minimization}
\label{sec:surface_evolver}

To numerically verify Eq.~\eqref{eq:contact_area}, we used the finite-element based software \emph{surface evolver} to minimize surface energy [Eq.~\eqref{eq:surface_energy}] by the gradient descent method~\cite{brakke1992surface}. Initialization and procedural-control scripts were implemented in Mathematica \cite{Mathematica} (notebook available at https://git.embl.de/dullwebe/dullweber2024). To numerically verify Eq.~\ref{eq:contact_area}, we computed the minimal energy configurations for a pair of droplets ($n=1$), a line of 7 droplets ($n=2$), a $5\times5$ lattice ($n=4$) and a $5\times5\times5$ droplet lattice ($n=6$) and measured the contact area of the central droplet. Bash scripts to rerun this analysis are provided at https://git.embl.de/dullwebe/dullweber2024.

\end{document}